%% file: main.tex
\documentclass[letterpaper,twocolumn,10pt]{article}
\usepackage{usenix2019_v3}
\usepackage[letterpaper, hmargin=1in, vmargin=1.2in]{geometry}
\usepackage[utf8]{inputenc}
\usepackage{todonotes}
\usepackage{xcolor}
\usepackage{pgfplots}
\pgfplotsset{compat=newest}
\definecolor{SnippetColor}{HTML}{EEEEEE}
\usepackage[frozencache=true,cachedir=minted-cache]{minted}
\setminted{breaklines}
\setminted{bgcolor=SnippetColor}
\newenvironment{code}{\captionsetup{type=listing}}{}
\usepackage{fancyvrb,fvextra}
\usepackage{multirow}
\usepackage{
	amsmath, 
	amsthm, 
	amssymb,
	fancyhdr, 
	setspace 
}
\usepackage{mathtools}
\usepackage[center,small,sc]{caption}
\usepackage{array}
\usepackage{subfig}
\usepackage[algo2e,linesnumbered,ruled]{algorithm2e}
\usepackage{algorithm}
\usepackage{algpseudocode}
\algdef{SE}[DOWHILE]{Do}{doWhile}{\algorithmicdo}[1]{\algorithmicwhile\ #1}%
\usepackage{blindtext}
\usepackage{colortbl}
\usepackage{afterpage}
\usepackage{comment}
\usepackage{lipsum}   % for filler text
\usepackage{mdframed}

\newmdtheoremenv{construction}{Construction}

\newtheorem{lemma}{Lemma}
\newtheorem{definition}{Definition}
\newtheorem{game}{Game}

\newminted[mycode]{bash}{}
\newcommand{\mycmd}[1]{\textcolor{blue}{#1}}

\newcommand{\rr}[2][]{%
\ifthenelse{\equal{#1}{}}{\todo[inline,backgroundcolor=green!25]{Reza: #2}}
    {\ifdef{\showDone}{\todo[inline,backgroundcolor=green!25]{Reza: #2 \newline \textcolor{red}{#1}}}{}}%
}

\newcommand{\mv}[2][]{%
\ifthenelse{\equal{#1}{}}{\todo[inline,backgroundcolor=red!25]{Mayank: #2}}
    {\ifdef{\showDone}{\todo[inline,backgroundcolor=red!25]{Mayank: #2 \newline \textcolor{blue}{#1}}}{}}%
}
\newcommand{\md}[2][]{%
\ifthenelse{\equal{#1}{}}{\todo[inline,backgroundcolor=teal!25]{Marten: #2}}
    {\ifdef{\showDone}{\todo[inline,backgroundcolor=teal!25]{Marten: #2 \newline \textcolor{red}{#1}}}{}}%
}

\hypersetup{
    colorlinks=true,
    linkcolor=blue,
    filecolor=magenta,      
    urlcolor=cyan,
}
\title{Recursive Augmented Fernet (RAF) Token \\ Alleviating the Pain of Stolen Tokens }

\begin{document}
\author{
  Reza Rahaeimehr, Augusta University, GA, USA, \texttt{rrahaeimehr@augusta.edu} \\
  Marten van Dijk, University of Connecticut, CT, USA,
  \texttt{vandijk@engr.uconn.edu}
}
\maketitle
% \tableofcontents
\newpage
\input{sections/abstract}
\input{sections/introduction.tex}

\input{sections/openstack.tex}
\input{sections/keystone.tex}

\input{sections/fernet.tex}

\input{sections/structure.tex}
\input{sections/proof.tex}

\input{sections/implementation.tex}

\input{sections/summary}

\bibliographystyle{IEEEtran}
\bibliography{main} 
%\ifCLASSOPTIONcaptionsoff
\newpage
%\fi
\appendix

\newpage
\onecolumn
\input{app/AppProof.tex}

\newpage
\section{Source Code}
\input{codes/create_vm}

\input{codes/create_vol}

\input{codes/howToExtractPayload.tex}
\input{codes/RaftLib.tex}
\input{codes/validateToken.tex}
\input{codes/expSet1.tex}
\input{codes/expSet2.tex}

\newpage

\end{document}

%% file: sections/abstract.tex
\begin{abstract}A robust authentication and authorization mechanism is imperative in modular system development, where modularity and modular thinking are pivotal. Traditional systems often employ identity modules responsible for authentication and token issuance. Tokens, representing user credentials, offer advantages such as reduced reliance on passwords, limited lifespan, and scoped access. Despite these benefits, the "bearer token" problem persists, leaving systems vulnerable to abuse if tokens are compromised. We propose a token-based authentication mechanism addressing modular systems' critical bearer token problem. The proposed mechanism includes a novel RAF (Recursive Augmented Fernet) token, a blacklist component, and a policy enforcer component. RAF tokens are one-time-use tokens, like tickets. They carry commands, and the receiver of an RAF token can issue new tokens using the received RAF token. The blacklist component guarantees an RAF token can not be approved more than once, and the policy enforcer checks the compatibility of commands carried by an RAF token. We introduce two variations of RAF tokens: User-tied RAF, offering simplicity and compatibility, and Fully-tied RAF, providing enhanced security through service-specific secret keys. We thoroughly discuss the security guarantees, technical definitions, and construction of RAF tokens backed by game-based proofs. We demonstrate a proof of concept in the context of OpenStack, involving modifications to Keystone and creating an RAFT library. The experimental results reveal minimal overhead in typical scenarios, establishing the practicality and effectiveness of RAF. Our experiments show that the RAF mechanism beats the idea of using short-life Fernet tokens while providing much better security. 
\end{abstract}

%% file: sections/introduction.tex
\section{Introduction}
Modularity and modular thinking and development models are the key to implementing an extensive system. For a modular system, a proper authentication and authorization mechanism is essential.
%Modern authentication and authorization mechanisms utilize the concept of tokens. 
Usually, there is a module, say the identity module, in a modular system that is in charge of authentication and authorization. Users present their credentials (usually a username and password) to the identity module and are granted tokens in exchange. Other modules only serve users who have valid tokens. Tokens are either %\textcolor{red}{??}  corrected
a random bit string like a UUID that points to an entry in an authentication and authorization database, or a crafted piece of data that carries some authentication and authorization information like Fernet and JWT tokens (see section \ref{sec:tokenFormats}).

Using token instead of username and password has three main advantages:
\begin{itemize}
    \item Users do not need to use their passwords for every system access. Consequently, the probability of leaking passwords gets lower. 
    \item Every token has an expiration time. Hence, a stolen token can only be abused for a limited time. 
    \item It is possible to limit the scope of a token and let the user ask for a less privileged token.  In this case, if an adversary steals the token, she will have limited access to the resources of the token owner. 
\end{itemize}
The above features allow a user to get a short-life scoped token from the identity module and delegate the authority to a third trusted party to accomplish a complicated task within the scope.  %\remhm{For example, in OpenStack, a user can ask the compute module to create a virtual machine (VM) on a specific network. The compute module is not in charge of networking. However, having the user token, the compute module can ask the networking module to add the VM to the desired network on behalf of the user. }
 \paragraph{Problem:} Although the ability to use a token and make requests on behalf of a user (authorization delegation) is a vital feature for a highly modular system, it is the source of many vulnerabilities \cite{hogan2019}. 
Adversaries can find bugs in large scale systems; With the current token mechanisms \cite{oauth,openid,jws}, if an adversary exploits some of these bugs and finds a way to obtain user tokens, she can do whatever the owners of the tokens can do. This problem is known as the ``bearer token'' problem \cite{hogan2019}. 

Scoping tokens down and reducing their lifetime are two ways that are used to reduce the impact of the bearer token problem \cite{tokenSolutions}. However, all the typical token types like UUID, PKI, Fernet, JWT \cite{jws} and ticketing systems like Kerberos \cite{KerberosForKeystone} have two main shortcomings:
\begin{itemize}
    \item First, in current mechanisms, the immediate impact of shortening the lifetime of tokens or reducing the scope of tokens is that users need to renew their tokens more often. In practice, if we remarkably shorten the lifetime of tokens and reduce the scope of tokens, the identity module turns into a bottleneck. 
    \item Second, no matter how much a system shortens the lifetime of tokens or limits the scope of tokens,  a stolen token can be very dangerous. This is because there is no relation between tokens and commands. For example in OpenStack, if you get a 1-second life token for deleting a volume in a project and the token is leaked to an adversary,  the adversary can delete all the volumes in the project during that 1 second.
\end{itemize}

% fixed \textcolor{red}{this sentence repeats a sentence from above} 
Considering this problem, system designers make a trade-off between system performance and security. For example, in OpenStack\cite{OpenStack}, the default token lifetime is one hour, and the smallest scope is a project\cite{TokenDefault}. In summary, despite  shortening the life time or reducing the scope of a token, an adversary who can corrupt a module and observe a token can do tremendous damage. 
Having a bug free system is not realistic, but reducing the effects of buggy code %\textcolor{red}{(in terms of corrupted modules and services)} 
is possible. Therefore, it is desirable to generalize the modularity security requirement that is proposed by Maleki et al. \cite{hogan2019} as follow:

\paragraph{Modularity Security Requirement:} For a modular system, if an adversary corrupts a module, other modules just do whatever requested to do by the users, nothing more. 

For an instance, in OpenStack, it is desired that, in the event an adversary compromises the Networking module, he should be unable to create or delete a VM. 

\paragraph{Our work:} This work provides the modularity security requirement for modular and distributed systems by developing the following parts:
\begin{enumerate}
    \item A novel token that we call Recursive Augmented Fernet (RAF) token on top of which commands can be added to tokens without requiring additional identity module interaction. 
    \item A blacklist structure that prevents replay attack. 
    \item A policy enforcer component that prevents malicious command execution. 
\end{enumerate}

We introduce two subtypes: \textit{User-tied RAF (Ut-RAF)} that is the best option for the systems that are currently using Fernet tokens and seeking for more secure solutions without the need for significant changes, and \textit{Fully-tied RAF (Ft-RAF)} that provides better security guarantees. Ft-RAF requires each module to have a secret key shared with  identity module. Hence, it needs a key distribution/renewal mechanism, which makes it harder to deploy compared to Ut-RAF. A summary of security analysis of Ut-RAF and Ft-RAF is given in section \ref{sec:RAFTsec}, and the details can be find in \ref{app:proof}.

We explain our proposal in the context of OpenStack to show its usability and performance in practice. OpenStack is one of the world's most successful open-source software that has been developed by tens of thousands of contributors around the world, offers a robust and smooth Infrastructure-as-a-Service (IaaS) cloud platform, and has been utilized by companies like RedHat, American Express,  IBM, AT\&T, Adobe, and Best Buy \cite{OpenStackSite}. % [ref].
%https://www.featuredcustomers.com/vendor/openstack/customers
Therefore, we firmly believe that if RAF tokens function effectively in OpenStack, they will easily integrate with other systems.

Our proposal has two important strength:

First, it allows reducing the lifetime or scope of tokens without interacting with identity module. It ties tokens to commands and in fact, limits the usage of tokens only to the commands issued by users and minimizes the impact of stolen tokens.

Second, it offers to the third-party Fernet-token-base system add-ons the choice to continue using Fernet and gradually upgrading to RAF (if the system policy continues accepting Fernet tokens). Note that, in our solution, the identity module still issues Fernet tokens.  

We have implemented a RAF library that allows issuing RAF tokens either based on Fernet or other RAF tokens. Also, we have modified the identity module of OpenStack to accept and validate RAF tokens. The experimental results show that our mechanism adds less than 1 percent overhead to the validation time of a token, i.e.,  less than a microsecond. %Usually, cloud commands like creating a server, migrating a server, and loading an image are very time-consuming tasks (in the order of several seconds). Hence, the overhead of our mechanism is negligible. Note that in most cases, a token validation time composes less than one percent of a user request's execution time. 
Also, our experiment showed that generating a RAF token from a long life Fernet token is significantly (more than 88 times in our experiments) faster compared to getting a new short lifetime Fernet token from the identity module.  
\paragraph{Threat Model:}
A service\footnote{In OpenStack, each module is referred to as a service. Therefore, in this article, we use the terms `module' and `service' interchangeably.} can be corrupted in two ways:
\begin{itemize}
    \item Partial corruption: where a service leaks tokens to the adversary, but it does what is supposed to do.
    \item Fully corruption: where a service is under control of the adversary and the adversary can do whatever he wants within the service.
\end{itemize}
We consider a strong adversary who can fully corrupt some services and observe all network communications except User-
identity module communications. Hence, user-identity module communications must be confidential and authenticated. % (secured by TLS and a Certificate)
In OpenStack, it is possible to secure User-Service communication by enabling TLS and having a certificate for each service. But, since service-to-service communication is not secure, we also assume User-Service communication is not secure. This allows us to cover a wider range of applications and provide  stronger security. 
\paragraph{Organization:}
We commence with an overview of the history and structure of OpenStack, elucidating the four popular token formats in Section \ref{sec:openstack}. Section \ref{sec:RAFTsec} provides details of our tokening mechanism, while Section \ref{sec:proof} delves into the security analysis of RAF. In Section \ref{sec:implementation}, we present a proof of concept for RAF and describe some of the results obtained from using RAF in our experiments. We conclude our work with a summary in Section \ref{sec:summary}.

%% file: sections/openstack.tex
\section{Background} \label{sec:background}
\subsection{OpenStack}\label{sec:openstack}
In 2010, Rackspace wanted to redesign its Cloud servers offering infrastructure, while NASA was interested in a similar project. Negotiation between the two teams led to a shared program called OpenStack. %RackSpace hosted the first OpenStack summit on July 2010 in Austin, Texas where the first OpenStack design was presented to more than 25 companies including AMD, Dell, Cloud.com, Intel, and Nebula. A few days after the summit, RackSpace and NASA with the contribution of more than 25 partners funded OpenStack aiming “to produce the ubiquitous Open Source Cloud Computing platform that will meet the needs of public and private clouds regardless of size, by being simple to implement and massively scalable.”\cite{OpenStackHistory} Three months later, on October first, 2010, OpenStack Austin was released. Austin includes only Nova and Swift projects. 
The OpenStack Foundation was established in September 2012, which has been in charge of OpenStack since then. In 2019, it had more than 100,000 community members from 187 countries who had collaborated to build one of the most significant open-source projects in the world \cite{OpenStackContributers}.

%The OpenStack Foundation updated the mission of OpenStack in February of 2016 as follows: ``to produce a ubiquitous Open Source Cloud Computing platform that is easy to use, simple to implement, interoperable between deployments, works well at all scales, and meets the needs of users and operators of both public and private clouds." \cite{OpenStack} In the updated mission statement, interoperability was added as a new interest, and satisfying the user needs was emphasized. 

%Since Austin, OpenStack has been developed and released around 6-month cycles. The name of the latest stable release (until September 2019) is Stein, and Train is under development.

OpenStack is a modular system divided into projects (services) at the highest level. Every project utilizes third-party plugins that must be open-source as well. Usually, for a use case, OpenStack supports several plugins that can be used interchangeably. The following projects are the core of OpenStack:
\begin{description}
\item[Keystone] implements Identity API and is in charge of authentication and authorization.
\item[Nova] is the heart of OpenStack and manages computation resources.
\item[Cinder] provides block storage (volumes) mainly used by virtual machines.
\item[Glance] is the imaging service, which maintains data assets like operating systems.
\item[Neutron] enables networking services. 
\item[Swift] provides efficient block storage and can be used independently. Hence, some companies take advantage of Swift for storing data. 
\item[Horizon] resembles the dashboard of OpenStack and provides a web interface for clients. Although Horizon is not the only way for interacting with OpenStack, it is the simplest way for clients if they want to manage the system manually. 
\end{description}
%For example, a resource provider can be a compute node, a shared storage pool, or an IP allocation pool. The placement service tracks the inventory and usage of each provider. For example, an instance created on a compute node may be a consumer of resources such as RAM and CPU from a compute node resource provider, disk from an external shared storage pool resource provider and IP addresses from an external IP pool resource provider.
In addition to Horizon, users can send their request to OpenStack via cURL \cite{curl} or OpenStack Client library. cURL allows a person to send an HTTP/HTTPS request. OpenStack Client library is a tool for developers and enables them to converts users' high-level commands to corresponding cURL commands. 

%As mentioned earlier, OpenStack started with two projects. The number of projects and the variety of services has been growing quickly. OpenStack Stein includes 42 projects, and Train extended to 45 projects. The functionality and responsibility of every service have been evolving. Interested readers are refered to the OpenStack official website to see the updated list of projects and details.  A person does not need to deploy all the services. Only Nova, Keystone, Glance, and Neutron are enough for having a productive OpenStack with basic functionalities.

\begin{figure}[ht]
	\includegraphics[width=\columnwidth]{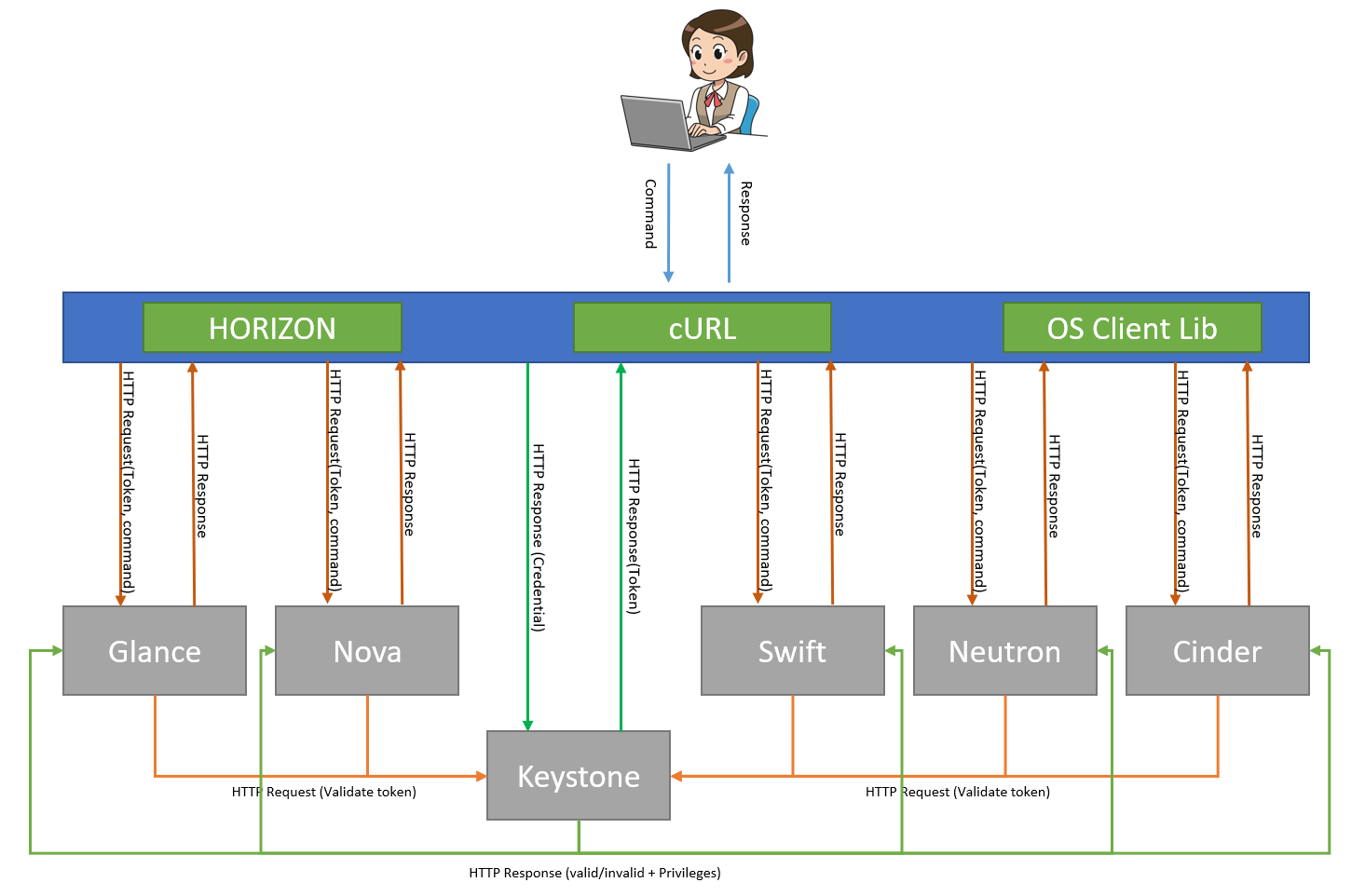}
    \centering
	\caption{The interaction between a user and the main OpenStack modules.}
	\label{fig:openstackinteraction}
\end{figure}

Each service implements some Representational State Transfer (REST) APIs, which are the standard way to communicate with a service in OpenStack. A REST API is a web service that is callable by HTTP/HTTPS requests. The OpenStack REST APIs accept parameters in JSON format. To access OpenStack, first, a user must get a token from Keystone by presenting his/her username and password. Each token has an expiration time, and the user needs to refresh his token periodically. The user must put the token in the header of all his requests to other modules. Whenever a service receives a request,  first of all, it checks the presence of a token in the header. Then, the service validates the token with Keystone. Sometimes a service needs to make a request to other services on behalf of the user using the received token. For example, whenever a user wants to create a virtual machine, in addition to the virtual machine specification (like the amount of RAM and number of virtual CPU), he must specify a network. In this case, the user sends his request to Nova. Nova builds the VM, but it needs to get in touch with Neutron to establish the connection between the VM and the network. Figure \ref{fig:openstackinteraction} shows the interaction between a user and some OpenStack modules as described above. 

%% file: sections/keystone.tex
\subsection{Token Formats}\label{sec:tokenFormats}
The four most common types of tokens that also have been used in OpenStack are as follows:
\paragraph{UUID}{
A UUID token is nothing more than a random string (32 characters in OpenStack) and carries no information. Consequently, to validate the issued tokens, the system needs to store them and relative meta-data in a database. Implementing the UUID token mechanism is straightforward. Its main disadvantage is the token database. In a practical deployment, a UUID token database grows fast, and after a while,  inserting and retrieving data from the database turns into a bottleneck for high throughput systems. In addition, the complexity of establishing multiple identity nodes and scalability are other concerns that convinced the OpenStack community to look for alternative token mechanisms.}
\paragraph{PKI and PKIz}{
PKI token was the first attempt to utilize Public Key Infrastructure in OpenStack. A PKI token carries some meaningful data like the owner's specification, privileges,  issue time and expiration time of the token, and catalog list signed by the private key of the identity service. Therefore,  PKI tokens are very long, and in practice can exceed 8k bytes, which does not fit in an HTTP header by default. The motivation was to minimize the need for database access in Keystone. At first glance, Keystone does not need to store PKI tokens. A PKI token can be verified by anyone who has the public key of Keystone. OpenStack services do not need to get in touch with Keystone to discover the privileges of a PKI token. However, in practice, this is not accurate: A service still needs to check a token with Keystone to see if it is not revoked; Keystone has to keep the revoked tokens in a database (or similar structure).

PKIz is the compressed version of PKI that was an effort to make PKI tokens small. Nevertheless, this technique did not help much, and PKIz tokens were only 10 percent smaller than PKI tokens in real usages. Because of all the disadvantages described above, the OpenStack community abandoned the PKI and PKIz tokens.}
\paragraph{Fernet}{In 2013, Heroku introduced its secure message transmission method that was called Fernet. They encrypted a message and concatenated the ciphertext with the HMAC of the ciphertext and called the resultant a Fernet token. In 2015, the Keystone core team adapted the Fernet token for OpenStack. OpenStack Kilo is the first release of OpenStack that supports Fernet token, and nowadays, it is the most used token format. 

In OpenStack, Fernet tokens contain a payload that is encrypted and symmetrically signed by Keystone's secret key.  The payload usually includes the expiration time of the token,  the Id of the owner, an Audit Id, and the scope of the token. Any Keystone node that has the secret key can validate a Fernet token and extracts the payload. The main advantage of Fernet tokens is that there is no need for a database (backend) for storing and validating them. These features made the Fernet format the most successful token format in OpenStack. }
\paragraph{JWS}{The OpenStack Stein adapts the JSON Web Signature (JWS) token format that is a subtype of JSON Web Token (JWT)\cite{jws}.
Each JWS token contains a header, a payload, and the signature of the token. The OpenStack uses ES256 JSON Web Algorithm with a private key for signing JWS tokens. Hence, who has the corresponding public key can verify a JWS token. We can say that the JWS format is the updated version of the PKI format.  Unlike PKI tokens, JWS tokens contain little data, and the payload of a JWS token is not encrypted.
Consequently, a JWS token is much shorter than a PKI token. However, it is still about twice as larger as a Fernet token. Like Fernet tokens, JWS tokens are ephemeral; this means we do not need to store them in a token database and replicate the database across all Keystone nodes. There is no apparent preference between Fernet and JWS.} %In some use cases, a person may consider the following as advantages of JWS over Fernet:
%\begin{enumerate}
%    \item Since we only need the corresponding public key to verify a JWS token, we can have some Keystone nodes that  are only in charge of verifying tokens, and we do not need to share the secret (private) between all of them.  
%    \item JWT is becoming a standard way for inter-system communication. Hence, using JWS token may facilitate having a hybrid cloud or federated could. 
%\end{enumerate}
%On the order hand, as mentioned earlier, the payload of a JWS token is not encrypted, and there are some concerns among the OpenStack community about revealing internal data to the public by using not encrypted JWS tokens. 

%% file: sections/fernet.tex
\subsection{Fernet Tokens} \label{sec:fernet}
The Fernet format is the base of our solution, and understanding how it works is essential. Hence, we describe Fernet in more detail here. 
%\subsubsection{Structure}
Keystone needs a Fernet key to generate Fernet tokens. A Fernet key is a base64 URL safe string composed of a signing key and an encryption key. If we decode a Fernet key, we will have a 32-byte array, which the first half (128 bits) is the signing key, and the second half is the encryption key \footnote{It is possible to use larger key. Since OpenStack uses a fresh key every 1 hour, the OpenStack community believes that a 128 bit key for encryption and 128 bit key for HMAC provide enough security.}. Fernet utilizes symmetric-key cryptography as follows:
\begin{enumerate}
    \item SHA256 HMAC for signing. 
    \item AES 128 in CBC mode with a 128-bit Initialization Vector(IV) for encrypting the payload of a token
\end{enumerate}
If we have a close look into a Fernet token, we can distinguish five parts:
\begin{description}
\item[Version:]
The first byte of every Fernet token shows the version of the token. The first version is denoted by 128 (0x80), and since, there is only one version of Fernet that has been introduced, we can say every Fernet tokens starts with the number 128. 
\item[Timestamp:]
The second part of a Fernet token is a 64-bit unsigned big-endian integer that represents the creation timestamp of the token in Linux time format.

\item[IV:]
IV is a 128-bit random string that is used as the initial vector of AES encryption. 

\item[Ciphertext:]
OpenStack Stein has ten different payload types including unscoped,     domain scoped, project scoped, trust scoped,  federated unscoped,     federated project scoped, federated domain scoped, OAuth scoped, and system scoped. Depending on the type, the data in the payload of a token varies. The payload of a token at least contains the type, the owner, the method that was used for authenticating the token owner, the expiration time, and the audit Id of the token. The length of a payload must be a multiple of 128 bits. Hence, OpenStack uses  PKCS \#7 v1.5 technique \cite{RFC} to pad the payload. Then, OpenStack encrypts the payload.
%ref=https://tools.ietf.org/html/rfc2315#section-10.3

\item[HMAC:]
The last part of a Fernet token is the HMAC of all the four previous parts, which is calculated by using SHA256 HMAC. This part is known as the signature of a Fernet token. 
\end{description}

%% file: sections/structure.tex
\section{Recursive Augmented Fernet (RAF) Mechanism} \label{sec:RAFTsec}
As mentioned earlier, in the current authentication mechanisms, whenever an adversary captures a user token, he can do all the things that the owner of the token can do (the bearer token problem). Recursive Augmented Fernet (RAF) mechanism is our solution for the bearer this problem. The solution includes: RAF, a Policy Enforcer(PE) component, and a Blacklist. This section explains the solution. The main idea is as follows:
\begin{itemize}
    \item Each user is granted a Fernet token from the identity module as like as common OpenStack authentication process, i.e., the user presents his credential to Keystone and Keystone returns a Fernet token after validating the credential.
    \item The user keeps the Fernet token secret and does not pass the token to other modules. For every API request, the user generates a new RAF token using the Fernet token. The RAF token contains enough information about the API and the user request, and it is unforgeable. The user puts the RAF token in the header of the API request instead of the Fernet token.
    \item Whenever a service receives an API request, it takes the RAF token and validates the RAF with the identity module. Comparing to the Fernet mechanism, if the RAF was valid, the identity module returns extra information about the target API and the user request. The PE verifies if the API request matches with the information returned by the identity module.
    \item Whenever a service needs to call another service on behalf of the user, it derives a new RAF token from the token it has received and uses the new token for its request.
    \item Each RAF token carries enough information about a user request. Using this information and a Blacklist, the identity module only verifies a RAF token pointing to the user request {\em once} for each service. 

\end{itemize}
\subsection{RAF Tokens} \label{sec:RAFT}
The core of our solution is the RAF token. To explain the RAF structure, first, we define the following terms:
\begin{description}
\item[Root Token:] is a Fernet token that a user obtains from Keystone and uses to generate RAFs. $T_0$ represents a root token.

\item[Base-RAF:] is a RAF token derived from a root  token by users. $T_1$ represents a base-RAF.

\item[Service-RAF:] is a RAF token that is derived from a base-RAF or another service-RAF. Service-RAFs are generated by the internal services and denoted by $T_i$ where $i>1$. 
\item[Parent token:] If a token $T_i$ is derived from $T_{i-1}$, we say $T_{i-1}$ is the parent of $T_i$.
\item[Child token:] If a token $T_i$ is derived from another token $T_{i-1}$, we say $T_i$ is a child of $T_{i-1}$.
\item[RAF digest:] In a big picture, a Fernet or RAF token  $T_i$ is a message $m_i$ concatenated with the HMAC of the message under a certain key. Hence, we can represent a token $T_i=m_i \parallel HMAC(m_i)$. In our solution, the $HMAC(m_i)$ is called the RAF digest of a token and denoted by $H_i$. Hence, we may write $T_i=m_i \parallel H_i$ where $H_i=HMAC(m_i)$.
\item[Parent Message:] If $T_i=m_i \parallel H_i$ was derived from the token $T_{i-1}=m_{i-1} \parallel H_{i-1}$, then $m_{i-1}$ is the parent message of $T_i$. 
\end{description}

\subsubsection{Structure}\label{raft_struct}
%As mentioned earlier, each RAFT token is derived from a Fernet or another RAFT token. Therefore, the root of every RAFT token is a Fernet token. A Fernet token is composed of some stuff $m_0$ concatenated with the HMAC of the stuff under the first half of a Fernet key. So, we can write a Fernet token $T_0=m_0 \parallel HMAC(m_0,key_{F1})$, where $key_{F1}$ is the first half of the Fernet. Considering the root token $T_0$, 
We define the following structure for a RAF token:
\begin{equation}
V \parallel L^{m}_{i-1} \parallel m_{i-1} \parallel E_i \parallel R_i \parallel C_i \parallel H_i,
\end{equation}
where
\begin{description}
\item[Version ($V$):] A byte shows the version of the token. We identify the first version by 0x91. This field enables a system to support multiple versions of RAFs in the future.
\item[Length-of-parent-message ($L^{m}_{i-1}$):] To extract the parent message of a token, we need to know its length. This field of a short integer (2 bytes) determines the length of the parent message in bytes. 

\item[Parent-message ($m_{i-1}$):] The authentication module needs to extract the parent message of a token to verify the token. 

\item[Expiration-Time ($E_i$):]
It is a 64-bit unsigned big-endian integer, which shows the expiration time of the token. It is calculated based on the number of seconds past January 1, 1970.

\item[Randomizer ($R_i$):] A 64 bit random number used to randomize a RAF token. Without this field, although it is unlikely, two separate RAF tokens can be identical if the same user simultaneously issues them using a single Fernet token for the same purpose with the same parameters.

\item[Command ($C_i$):] This is the command that we want to execute using the token. 
\item[HMAC ($H_i$):]
This field is a 256-bit SHA256 HMAC that is used for authenticating the content of a token. The signing key of the HMAC and the fields included in the HMAC varies based on the token type explained in the following section.
\end{description}

\subsubsection{Variations} \label{sec_variations}
\begin{figure}[ht]
  \centering
  \subfloat[Fully-tied RAF.]{\includegraphics[width=0.45\textwidth]{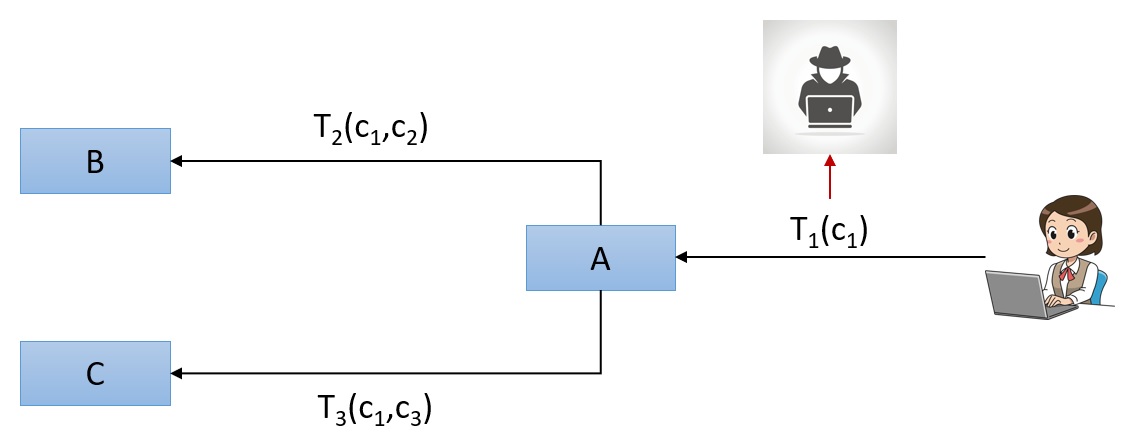}\label{pic:fully-tied}}
  \hfill
  \subfloat[User-tied RAF.]{\includegraphics[width=0.45\textwidth]{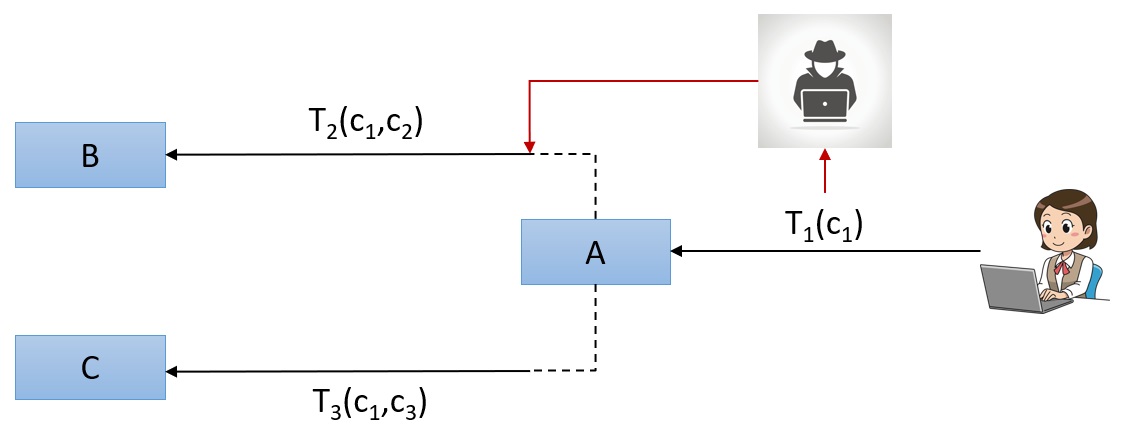}\label{pic:user-tied}}
  \caption{The effect of a leaked token in two types of RAF.}
  \label{fig:effectOfLeakedToken}
\end{figure}

We introduce two types of RAFT: 1- User-tied RAFT (UT-RAFT) 2- Fully-tied RAFT (FT-RAFT). The structure of both tokens is the same. But, they use different signing keys and includes different fields in HMAC as follows:
\begin{description}
\item[User-tied RAF:]
In this type, to derive a token from a parent token (received token), we take the first 128 bits of the parent key and use it for signing the new token. The HMAC field is the HMAC of all the fields prior to it in the token. This approach guarantees the authenticity of the user command carried in a RAF token. 
\item[Fully-tied RAF:]
In this type, the process of issuing a new RAF token for a user is the same as User-tied version. But, each service has a long-term symmetric key shared with the identity module and uses this key for signing tokens. The HMAC field is the HMAC of all the fields prior to it in the token plus the parent key of the token. Consequently, the RAF token is tied to the command and the service, and the identity module can verify the authenticity of the user and services who have contributed in a RAF token. 
\end{description} 
Figure \ref{fig:effectOfLeakedToken} shows the effect of a leaked token in two types of RAF with an example. In the example, the user issues the token $T_1$ for her command (request) $C_1$ and sends $T_1$ to the service $A$. To accomplish $C_1$, $A$ needs to send the $C_2$ and $C_3$ commands to $B$ and $C$, respectively. As shown in the picture, $T_1$ is leaked to the adversary. In the Fully-tied RAF, the adversary can not forge any valid RAF token using $T_1$, but in the User-tied RAF, the adversary can forge several valid RAF tokens. However, the point is that all of them carry $C_1$, and if the adversary put a command that does not align with $C_1$, the PE component (described later in this chapter) will reject the token.

\noindent The advantage of the fully tied token over the user tied token is that:
\begin{itemize}
    \item It guarantees stronger security (see section \ref{sec:proof}).
    \item It is not (computationaly) possible to forge a RAF token on behalf of a module unless the module was corrupted. 
\end{itemize}
However, since in fully tied RAF, each module must have a secret key shared with the identity module, we need to adopt a proper key distribution and key renewal mechanism, which in turn it induces extra efforts. 
While we possess the knowledge of how to perform key renewal \cite{kundu2019automatic,white2018client,adefarati2016integration}, the challenge lies in determining an appropriate key renewal and management strategy for large communities like OpenStack, which entails reaching a consensus within the OpenStack community regarding the security posture that the key renewal methodology should adopt. The user-tied RAF does not require such discussion. The fully-tied RAF needs an adaptation/extension of a/current key renewal practice. 

\subsubsection{Generation} \label{sec:generating}
% done \textcolor{red}{Here, and earlier I changed 'generating' to 'generation'. Similarly, we write verification and not verifying as subsection headings etc. Can you change this elsewhere?}
Algorithms \ref{alg:GenRAFTUser} and \ref{alg:GenURAFTService} show the steps for issuing a user-tied RAF token, and algorithms \ref{alg:GenRAFTUser} and \ref{alg:GenFRAFTService} represent the steps for issuing a fully-tied RAF token. In both RAF variants, users run the same algorithm.
As shown in the algorithms, given a parent token $Token=Pm\parallel Pkey$, a command $Cmd$, the life time $LifeTime$, and the service keys $KeySet$ (for Fully-tied RAF), we  follow the underneath steps to issue a RAF token:
\begin{enumerate}
    \item Unpack the parent token and get the parent message and parent key.
    \item Choose the proper signing key 
    \item Put the value 0x91 in the version field.
    \item Set the Length of the parent message field.
    \item Set the parent message.
    \item Add the token life time to the current time and put the result in the expiration-time field.
    \item Choose an eight-byte random number for the randomizer field.
    \item Set the command.
    \item For user-tied token: Compute the HMAC of all above fields using the parent-key.
    \item For fully-tied token: Compute the HMAC of all above fields plus the parent key using the secret key of the service
    \item Base64url encode the entire token.
\end{enumerate}
\begin{algorithm}
\SetKwProg{Fn}{Function}{}{}
\Fn{UserIssue(Key, Pm, Cmd, LifeTime)}
{
 $SignKey = HMAC(Key, Pm)$ \\
 $V = 0x91$\\
 $L=Length(Pm)$\\
 $E=GetCurrentTime()+LifeTime$\\
 $R \leftarrow Gen(1^{128})$\\
 $Payload = V \parallel L \parallel Pm \parallel E \parallel R \parallel Cmd$ \\
 $Signature = HMAC(SignKey,Payload) $ \\
 $Token = base64.Encode(Payload \parallel Signature)$ \\
 \textbf{return} $Token$\\
 }
 \bf{End Function}\\
  \caption{Pseudocode of Issuing a user-tied/fully-tied RAF token by a User.}
  \label{alg:GenRAFTUser}
\end{algorithm}

\begin{algorithm}
\SetKwProg{Fn}{Function}{}{}
\Fn{ServiceIssue(Key, Token, Cmd, LifeTime)}
{
 \eIf{$Verify(Key,Token)=1$}
 {
 $(Pm,PKey) = Unpack(Token)$ \\
 $V = 0x91$\\
 $L=Length(Pm)$\\
 $E=GetCurrentTime()+LifeTime$\\
 $R \leftarrow Gen(1^{128})$\\
 $Payload = V \parallel L \parallel Pm \parallel E \parallel R \parallel Cmd$ \\
 $Signature = HMAC(PKey,Payload) $ \\
 $Token = base64.Encode(Payload \parallel Signature)$ \\
 \textbf{return} $Token$\\
 }
 {
  \textbf{return} $nothing$
 }
 }
 \bf{End Function}\\
  \caption{Pseudocode of Issuing a user-tied RAF token by Services. }
  \label{alg:GenURAFTService}
\end{algorithm}

\begin{algorithm}
\SetKwProg{Fn}{Function}{}{}
\Fn{ServiceIssue(i,KeySet, Token, Cmd, LifeTime)}
{
\eIf{Verify($k_0 \in KeySet$,Token)=1}
{
 $Pm,PKey \leftarrow Unpack(Token)$ \\
 $SignKey=k_i \in KeySet$\\
 $V = 0x91$\\
 $L=Length(Pm)$\\
 $E=GetCurrentTime()+LifeTime$\\
 $R \leftarrow Gen(1^{128})$\\
 $Payload = V \parallel L \parallel Pm \parallel E \parallel R \parallel Cmd$ \\
 $ Signature = HMAC(SignKey,Payload\parallel PKey)  $\\
 $Token = base64.Encode(Payload \parallel Signature)$ \\
 \textbf{return} $Token$\\
 }
 {
 \textbf{return} $nothing$
 }
 }
 \bf{End Function}\\
  \caption{Pseudocode of Issuing a fully-tied RAF token by services. }
  \label{alg:GenFRAFTService}
\end{algorithm}

In section \ref{sec:implementation}, we show the implementation of algorithms \ref{alg:GenRAFTUser} and \ref{alg:GenURAFTService} in Python code.

\subsubsection{Verification}
Algorithm \ref{alg:verRAFTUser} shows pseudocode of a user-tied RAF token Verification. Given the secret key of Keystone $k_{keystone}$ and a token, to verify that the token is a valid user tied token and recover all the commands embedded in the token, perform the following steps, in order:
\begin{enumerate}
    \item base64url decode the token.
    \item If the first byte of the token is not 0x91, then the token is not valid. Therefore, raise an exception error.
    \item Unpack the token.
    \item Check the expiration time of the token in $m_i$. 
    \item Using the length-of-parent-message field (second and third bytes), retrieve $m_{i-1}$. If $m_{i-1}$ is a RAF-message (i.e. its first byte is 0x91), check the expiration time of $m_{i-1}$ and retrieve $m_{i-2}$ from $m_{i-1}$. Recursively continue this step until you get $m_0$ ,which must be a Fernet-message (i.e. its first byte must be 0x80). 
    \item Calculate $key_0 = HMAC (k_{keystone}, m_0)$
    \item Having $key_0$, backtrack step 5 and calculate all parent keys $\{key_1,\cdots\,key_{i-1}\}$ computing $key_j = HMAC(key_{j-1},m_j)$.
    \item Having $key_{i-1}$, recalculate the HMAC of the given token. If the calculated HMAC is equal with the HMAC of the given token, then verify the original Fernet token $m_0 \parallel key_0$. 
    \item Extract the Fernet message from $m_0$ and all commands from $\{m_1,\cdots,m_i\}$ and return them.
\end{enumerate}

\begin{algorithm}
\SetAlgoLined
\SetKwProg{Fn}{Function}{}{}
\Fn{Verify(Key, Token)}
{
 $Token = base64.decode(Token)$ \\
% $V \parallel L \parallel Pm \parallel E \parallel R \parallel Cmd \parallel Tag \leftarrow Unpack(Token)$ \\
 $V, L, Pm, E, R, Cmd, Tag \leftarrow Unpack(Token)$ \\
 \eIf{$V = 0x91$}
 {
    $CheckExpirationTime(E)$ \\
    $Cmds,PKey = ValidateParent(Key,Pm)$ \\
    $AccomulateCommands(Cmds,Cmd)$\\
    \eIf{$HMAC(PKey, V \parallel L \parallel Pm \parallel E \parallel R \parallel Cmd)=Tag$}
    {\textbf{return} $Valid,Cmds$}
    {\textbf{return} $Invalid$}
 }
 {
    \textbf{return} Invalid;
 }
 }
 \bf{End Function}\\
 \Fn{ValidateParent(Key,Message)}
{
     \eIf{First byte of Message is $0x91$}
     {
        $(V, L, Pm, E, R, Cmd \leftarrow Unpack(Message)$ \\
        $CheckExpirationTime(E)$ \\
        $Cmds,PKey = ValidateParent(Key,Pm)$ \\
        $AccomulateCommands(Cmds,Cmd)$\\
        $Tag = HMAC(PKey, V \parallel L \parallel Pm \parallel E \parallel R \parallel Cmd)$ \\
        \textbf{return} $Cmds,Tag$
     }
     {
        \textbf{return} $"",HMAC(Key, Message)$
     }
 }
 \bf{End Function}\\

  \caption{Pseudocode of a user-tied RAF token Verification}
  \label{alg:verRAFTUser}
\end{algorithm}

Verification of a fully-tied RAF token is slightly different from Verification a user-tied token. In fully-tied RAF, each service also has a secret key, and service uses its secret key to sign tokens. Algorithm \ref{alg:verRAFTFull} is the pseudocode for a fully-tied token verification. Given all the secret keys of the authentication module and other services $KeySet =\{k_0,\cdots,k_n\}$ and a token, to verify that the token is a valid fully tied token and recover all the commands embedded in the token, perform the following steps, in order:
\begin{enumerate}
    \item base64url decode the token.
    \item If the first byte of the token is not 0x91, then the token is not valid. Therefore, raise an exception error.
    \item Unpack the token.
    \item Check the expiration time of the token in $m_i$. 
    \item Using the length-of-parent-message field (second and third bytes), retrieve $m_{i-1}$. If $m_{i-1}$ is a RAF-message (i.e. its first byte is 0x91), check the expiration time of $m_{i-1}$ and retrieve $m_{i-2}$ from $m_{i-1}$. Recursively continue this step until you get $m_0$ ,which must be a Fernet-message (i.e. its first byte must be 0x80). 
    \item Calculate $tag_0 = HMAC(k_0,m_0)$, where $k_0$ is the secret key of Keystone.
    \item Having $tag_0$, backtrack step 5 and calculate all the tags of ancestor tokens $\{tag_1,\cdots\,tag_{i-1}\}$ computing $tag_j = HMAC(k_j,m_j \parallel tag_{j-1})$.
    \item Having $tag_{i-1}$, recalculate the HMAC of the given token. If $HMAC(k_i,m_i \parallel tag_{i-1}) = tag_i$, then verify the original Fernet token $m_0 \parallel tag_0$. 
    \item Extract the Fernet message from $m_0$ and all commands from $\{m_1,\cdots,m_i\}$ and return them.
\end{enumerate}

\begin{algorithm}
\SetKwProg{Fn}{Function}{}{}
\Fn{Verify(KeySet, Token)}
{
 $Token = base64.decode(Token)$ \\
 $(V \parallel L \parallel Pm \parallel E \parallel R \parallel Cmd \parallel Tag) \leftarrow Unpack(Token)$ \\
 \eIf{$V = 0x91$}
 {
    $CheckExpirationTime(E)$ \\
    $k_i = FindServiceKey(Cmd)$ \\
    $Cmds,PTag = ValidateParent(KeySet,Pm)$ \\
    $AccomulateCommands(Cmds,Cmd)$\\
    \eIf{$HMAC(k_i, V \parallel L \parallel Pm \parallel E \parallel R \parallel Cmd \parallel PTag)=Tag$}
    {return $Valid,Cmds$}
    {return $Invalid$}
 }
 {
    return Invalid;
 }
 }
 \bf{End Function}\\
 \Fn{ValidateParent(KeySet,Message)}
{
     \eIf{First byte of Message is $0x91$}
     {
        $(V \parallel L \parallel Pm \parallel E \parallel R \parallel Cmd) \leftarrow Unpack(Message)$ \\
        $CheckExpirationTime(E)$ \\
        $k_i = FindServiceKey(Cmd)$ \\
        $Cmds,PTag = ValidateParent(KeySet,Pm)$ \\
        $AccomulateCommands(Cmds,Cmd)$\\
        $Tag = HMAC(k_i, V \parallel L \parallel Pm \parallel E \parallel R \parallel Cmd \parallel PTag)$ \\
        return $Cmds,Tag$\\
     }
     {
        return $"",HMAC(k_0, Message)$
     }
 }
 \bf{End Function}\\

  \caption{Pseudocode of a fully-tied RAF token verification }
  \label{alg:verRAFTFull}
\end{algorithm}

\subsection{Blacklist} \label{sec:backlist}
The Blacklist not only prevents replay attacks, it also guarantees that each module serves at most once for every user command. Every RAF token is either a base-RAF token or has been derived from a base-RAF token. Hence, we can say every valid RAF token has a unique base-RAF token. We want each base-RAF token to be a permit for only one task in the system. For example, if a user generates a base-RAF for accessing an image, we should guarantee that the image will be accessed at most once using this user-RAF. To do so, we propose to implement a blacklist mechanism in the token validation process. Each entry in the blacklist is a tuple $(B, S, E)$ that is added to the list upon a successful token validation where:
\begin{description}
\item[$S$:] is the service who asked the authentication module to validate the token $T$.
\item[$B$:] is the base-RAF of $T$.
\item[$E$:] is the expiration time of $B$. 
\end{description}
When $S$ asks the authentication module to validate $T$, it extracts $B$ out of $T$.  It also extracts the expiration time $E$ out of $B$. Since for executing user commends each module needed at most once, Keystone only validates a token when its corresponding tuple $(B, S, E)$ was not in the blacklist. Keystone periodically cleans up the blacklist and keeps only the tuples that have not been expired. A tuple $(B, S, E)$ is expired when $E$ is less than system time. Since the lifetime of a RAF token is usually very short, a lightweight memcache structure works fine. 
\subsection{Policy Enforcer} \label{sec:PolicyEnforcer}
\begin{figure}[ht]
	\includegraphics[width=\columnwidth]{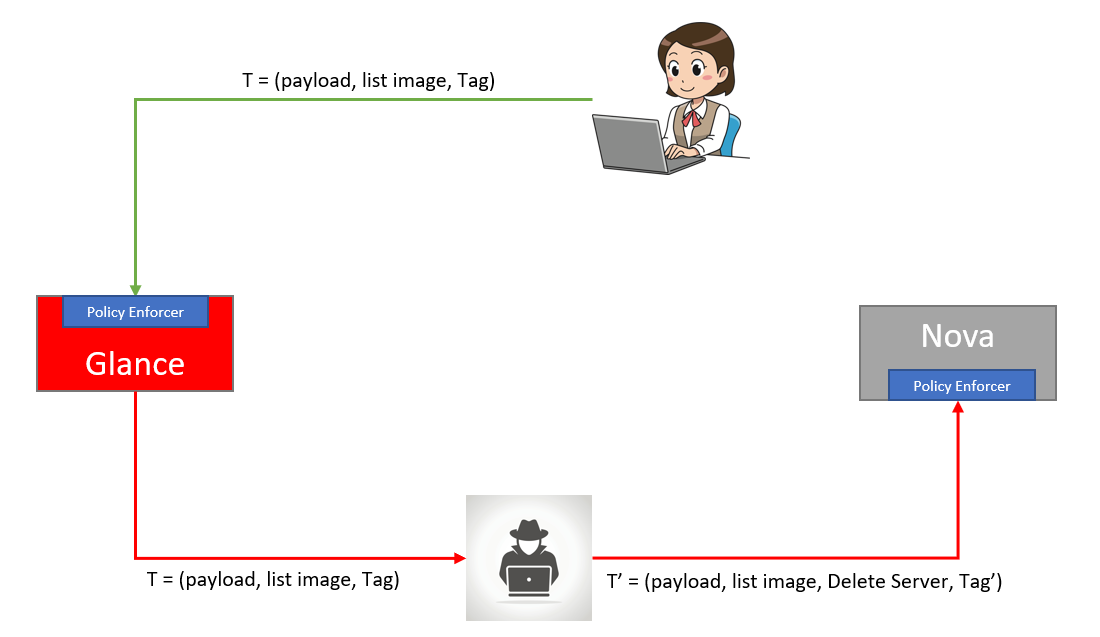}
    \centering
	\caption{Demonstration of the usage of Policy Enforcer}
	\label{fig:policyenforcer}
\end{figure}

If the adversary obtains a user tied RAF token, he/she can produce children of the token. Also, in fully tied RAF, if the adversary knows the secret key of a module (corrupts the module), he/she can generate children for the tokens sent to the module. RAF (either user tied or fully tied) excludes an adversary to produce a token without knowledge of one of its parents/ancestors.  The policy enforcer is used to make sure that the command embedded in a RAF token is consistent and obeys a proper execution flow (specified by a policy). This means that even though an adversary can create child tokens, he/she can only do so using commands that fit the policy enforcer. Figure \ref{fig:policyenforcer} demonstrates the idea using OpenStack. Here, the user creates a token $T$ for the ``list image" command and sends it to Glance, which is compromised by the adversary. Hence, the adversary can create a child token like $T'$. If the adversary adds an arbitrary command like ``delete a server" to the new token, and sends the token to Nova, since the "delete server" command does not relate the ``list image" command, the policy enforcer of Nova will reject the request. 
The policy enforcer completes our solution.

%% file: sections/proof.tex
\section{Security Guarantees}\label{sec:proof}

%\textcolor{red}{Start with some intuition about what you are going to do: an outline, may be introduce subsubsections?}
%%%%%%%%%%%%%%%%%%
%%%%%%%%%%%%%%%%%%

The aim of RAFT tokens is to prevent an adversary from sending a request to a modular system on behalf of the user.  %Both type of RAFT tokens prevents the adversary from impersonating a legitimate user. However, the fully tied RAFT token provides additional security, that is, it prevents the adversary to send a token on behalf of a service. In other word, if the adversary was able to get access to a fully tied token, s/he is not able to send a request on behalf of a service to another service.
At a high level, a RAF token is an algorithm that gets a message as an input and outputs a token. 
Security is formulated by requiring that no adversary can generate a valid token on any new message. For user tied RAF token, a new message is considered to be any message that the message itself or its ancestors was not previously sent by a user or other modules. For fully tied RAF, only the message itself should not be previously sent by user or other modules.

In this section we want to define the security that each type of RAF mechanism provides. In order to do this we take the following steps:
 \begin{itemize}
     \item First, we present the technical definition of what a user/fully tied RAF is.
     \item Next, we define an experiment as a game between the RAF mechanism and the adversary. The experiment shows how an adversary interacts with the mechanism and in which circumstances the adversary may successfully forge a RAF token and win the game.
     \item Using this game and the technical definition of RAF mechanism, we provide the definition of the RAF token security. The definition states that no PPT adversary should win the game with non-negligible probability.  
     \item Finally, we construct a secure user/fully tied RAF using the algorithms defined in section \ref{sec:generating} and define a lemma which indicates that our construction is secure and unforgeable against a chosen-message attack.
 \end{itemize} 

The RAF token comes into two flavors, user tied and fully tied RAF token. In section \ref{sec:AnalysisUserRAFT} we argue about the security guarantees of user tied RAF mechanism while in section \ref{sec:AnalysisFullyRAFT} we focus on the security of fully tied RAF.
\subsection{The Security Analysis of User Tied RAF}\label{sec:AnalysisUserRAFT}
A user tied RAF token prevents the adversary from impersonating a legitimate user. Before defining the security features of this mechanism, we first provide a technical definition of user tied RAF. Technically, a user tied RAF is made up of four algorithms $Gen$, $UserIssue$, $ServiceIssue$, and $Ver$. The algorithm $Gen$ generates a secret key; we assume that upon input the security parameter $\lambda$, the algorithm $Gen$ outputs a uniformly distributed string of length $\lambda$. The algorithms $UserIssue$ and $ServiceIssue$ receive a message/token ($p$/$r$), a command $c$ and a key $k$ as an input and generate a token. $UserIssue$ is used by the user dashboard while $ServiceIssue$ is used by the other services. Finally, the algorithm $Ver$ receives a token $r$ and a key $k$ and outputs either 1 (meaning valid) or 0 (meaning invalid). In the following we have the formal definition:

\begin{definition}\label{def:user-tiedToken}
A User-tied RAF token is a quadruple $\Pi = (Gen, UserIssue, ServiceIssue, Ver)$ of ppt algorithms where:
\begin{itemize}
    \item $k\leftarrow Gen(1^\lambda)$ , where $Gen$ generates a key $k$ with security parameter $\lambda$
    \item $r_0=(m_0,t_0) \leftarrow UserIssue(k,p,c_0)$, where $UserIssue$ generates a token $r_0$ for the root payload $p$ and the command $c_0$ with the key $k$. 
    \item $r_j=(m_j,t_j) \leftarrow ServiceIssue(k,r_{j-1},c_j)$, where if $r_{j-1}$ is internally verified by $Ver(k,r_{j-1})$ then $ServiceIssue$ generates a token $r_j$ for the input token $r_{j-1}$ and the command $c_j$ with the key $k$; Otherwise it returns nothing. 
    
    \item $b \leftarrow Ver(k,r_n=(m_n,t_n))$, where $Ver$ takes the key $k$ and a token $r_n$. It outputs $b$, with $b=1$ meaning the token is valid and $b=0$ meaning the token is not valid.

    \item It is called correct if for every $k$ output by $Gen(1^\lambda)$ and every token $r$ generated by $UserIssue$ or $ServiceIssue$; it holds $Ver(k,r)=1$.
\end{itemize}
\end{definition}

\noindent A user-tied RAF token must provide the following properties:
\begin{itemize}
    \item Represents a user-command.
    \item Enables a module to extend a token with a command.
    \item Prevents an adversary to forge any token except some descendants of the eavesdropped tokens.
\end{itemize}
A descendant of a token contains all the commands in the token.  If the adversary adds an irrelevant command to a child token of an eavesdropped token, it will be rejected by the policy enforcer (see section \ref{sec:PolicyEnforcer}) of other modules. Hence, the ability to forge a child token is tolerated. %done \textcolor{red}{also refer back to the policy enforcer subsection}

Now, we define the following game for a PPT adversary ${\cal A}$, the security parameter $\lambda$, and a user-tied RAF token $\Pi = (Gen, UserIssue, ServiceIssue, Ver)$:

\begin{game} \textbf{Forge a User Tied RAF token $ForgeUTT_{{\cal A},\Pi } (\lambda)$:}\label{gm:UTT}
\begin{enumerate}
    \item $Gen(1^\lambda)$ generates a key $k$
    \item The adversary $\cal A$ is given input $1^\lambda$ and oracle access to $Ver(k,.)$, $UserIssue(k,.,.)$, and $ServiceIssue(k,.,.)$ and outputs $r=(m,t)$. Let $\Phi$ represent the set of all tokens that $\cal A$ has queried.
    \item $\cal A$ wins if and only if (1) $Ver(k,r=(m,t))=1$ and (2) $r$ and any of its ancestors are not in $\Phi$.%its ancestors .
    \item The experiment returns 1 if the adversary wins the game, otherwise it returns 0.
\end{enumerate}
\end{game}
The definition of game \ref{gm:UTT} covers our goal. It ties a token to the user-command and allows a service to extend the token with a new command. Since the adversary has oracle access to $UserIssue$ and $ServiceIssue$, it covers the situation that the adversary eavesdrops some tokens. It states that no PPT adversary should be able to generate a valid user-tied RAF token unless he has seen at least one of its ancestors. This guarantees that the adversary cannot forge a token with a fake user command.

\begin{definition} \label{def:userSecure}
A user tied token $\Pi = (Gen, UserIssue, ServiceIssue, Ver)$ is unforgeable under a chosen-message attack, or just secure, if for all PPT adversaries $\cal A$ there exists a negligible
function\footnote{A function {$\displaystyle f(x):\mathbb {N} \to \mathbb {R} $} is negligible, if for every positive polynomial $poly(·)$ there exists an integer $N_{poly} > 0$ such that for all $x > N_{poly}$; it holds
    {$\displaystyle |f(x)|<{\frac {1}{{\text{poly}}(x)}}$}.} \textbf{negl} such that:
\begin{equation} 
    \Pr[ForgeUTT_{{\cal A}, \Pi} (\lambda)=1]\leq negl(\lambda).
\end{equation}
\end{definition}
Using definition \ref{def:user-tiedToken} and algorithms \ref{alg:GenRAFTUser}, \ref{alg:GenURAFTService}, and \ref{alg:verRAFTUser} we construct a user tied RAF token. We then prove that our construction is a secure user tied RAF.

\begin{definition}\label{def:HASH}
We say HMAC is secure and indistinguishable from a random function, %, i.e. the probability of distinguishing the value of an HMAC on an unobserved point from a random number is negligible for a polynomial time distinguisher $D$. 
if for all probabilistic polynomial-time distinguishers D, there exists a negligible
function $\alpha$ such that:
\begin{align}\label{eq:distinguish1}
|\Pr[D^{HMAC(k,.)}(\lambda)=1]- \Pr[D^{f(.)}(\lambda)=1]| \leq \alpha(\lambda)    
\end{align}
where $f(.)$ is a random function and $\alpha(\lambda)$ is a negligible function in $\lambda$. 
\end{definition}

\begin{lemma}
Assume that the HMAC used in the Construction \ref{con:RAFT} is secure and indistinguishable from a random function. Then, Construction \ref{con:RAFT} is a secure user tied token that is unforgable under chosen message attacks.
\label{lem:UTRaft1}
\end{lemma}

\begin{construction}\textbf{User Tied RAF} \label{con:RAFT}

Let HMAC be indistinguishable from a random number. Define $RAFT=(Gen, UserIssue, ServiceIssue, Ver)$ as follows:
\begin{itemize}
    \item \textbf{$Gen(1^\lambda)$}: Upon input $1^\lambda$, choose $k \leftarrow \{0,1\}^{\lambda}$
    \item \textbf{$UserIssue(k,p,c)$}: Upon input key $k$ , payload $p$ and the command $c$ compute $r$ by running Algorithm \ref{alg:GenRAFTUser}.
    \item \textbf{$ServiceIssue(k,r',c')$}: Upon input key $k \in \{0,1\}^{\lambda}$, token $r'$ and the command $c'$ compute $r$ by running Algorithm \ref{alg:GenURAFTService}.
    \item \textbf{$Ver(k,r=(m,t))$}: Upon input key $k \in \{0,1\}^{\lambda}$, token $r$ compute $b$ by running Algorithm \ref{alg:verRAFTUser}.
\end{itemize}
\end{construction}
%%%%%%%%%%%%%%%%%%%%%%%%%%%%%%%%%%%%%%%%%%
%%%%%%%%%%%%%%%  BEGIN  %%%%%%%%%%%%%%%%%%
%%%%%%%%%%%%%%%%%%%%%%%%%%%%%%%%%%%%%%%%%%

In this section we only present an informal sketch of the proof of this lemma. We defer to Appendix \ref{app:proof} for the full proof.

\begin{proof}[Proof sketch]
Our argument proceeds via a reduction approach. That is, we show how an adversary breaking construction \ref{con:RAFT} can be used as a sub routine to violate the assumption that \textit{"HMAC is secure and indistinguishable from a random function."}

To be more precise, we first assume, by contradiction, that a polynomial-time adversary $\cal A$
manages to forge a user tied $RAF$ (breaks construction \ref{con:RAFT}), i.e. with a non-negligible probability $\epsilon$ s/he is able to win the game \ref{gm:UTT}. Next, we build a new scheme, $RAF'$ to be the same as user tied $RAF$ except that instead of using the $HMAC$ for key $k$ in its algorithms, i.e. in $UserIssue$ and $Ver$, it uses a random function. We show that this implies the existence of a polynomial-time algorithm, called distinguisher $\cal D$, that can distinguish the $HMAC$ from a random one with advantage $O(\varepsilon)$. This will then imply that $\varepsilon$ must be negligible; otherwise it will contradict the assumption that $HMAC$ is secure.

More precisely, we show that the user tied $RAF$ is secure and the upper bound on the probability of forging RAF is about a factor $n$ times $\epsilon$, where $n$ is the number of commands added in the forged RAFT token. This upper bound makes sense: Let us consider that a token can at most contain $n$ commands; we can imagine these commands leading to a HMAC chain. In the proof methodology, the attacker $\cal A$ can be successful in impersonating any spot in the chain by using the command $c_i$. This means that there are $n$ possibilities and thus the upper bound is a factor of $n$. By observing that $n=poly(\lambda)$ for a PPT adversary $\cal A$ the proof is completed.
\end{proof}

%\textcolor{red}{Let's put the proofs in an appendix: In this section you can focus on explaining the intuition behind the lemma(s) -- it tells us that is is secure and that you pay some price in the reduction, i.e., a factor $n$ etc. Explain why this makes sense: in the proof methodology the attacker can be successful in impersonating any spot in the chain by using the command $c_i$ -- if a chain is of length $n$, this will mean $n$ possibilities and gives the factor $n$}

%\textcolor{red}{Be precise. We have an exact formula: $(n+1)\cdot \alpha(\lambda) + 1/(2^\lambda - C)$ (notice that $\alpha(\lambda)$ does not depend necessarily need to depend on $C$, but $\beta(\lambda)$ has a much more direct dependence on $C$ as described by the proof; you can introduce e.g. the notation $\alpha_C(\lambda)$). You should explain this. Better is not to talk about $f(\lambda)$, usually $\lambda$ means that the domain has $\lambda$ bits and hence there are $2^\lambda$ choices. Also $f'(\lambda)$ looks like a derivative. The oracle O is not yet defined.}
\textit{Discussion.}
Lets assume that the HMAC which is used in the implementation has $\alpha(\lambda)=\frac{1}{2^{\lambda}}$, then the lemma gives a concrete security statement with $\epsilon (\lambda) \leq (n+1)\cdot \frac{1}{2^{\lambda}}+\frac{1}{2^{\lambda}-C}$ where $n$ is the number of modules in the system and $C=poly(\lambda)$ is the adversary's maximum number of queries. Note that because of the blacklist component, $n$ can not be more than the number of modules in a system.

We introduce the notation $\alpha_C(\lambda)= \frac{1}{2^{\lambda}-C}$. This implies that $\alpha(\lambda)<\alpha_C(\lambda)$. 
In the case that the adversary has made $C$ queries to the HMAC(k,.) vs f(.) oracle, then we can say, $\epsilon (\lambda) \leq (n+2)\cdot \alpha_C(\lambda)=(n+2)\cdot \frac{1}{2^{\lambda}-C}$. 
%where $C=poly(\lambda)$ is the adversary's maximum number of queries. 
If key $k$ does not change for a long time, this implies that the adversary can make more queries to the oracle $\cal O$ ($C$ gets too large) and therefore, $\epsilon$ gets too large, i.e. the adversary's chance of winning the game increases. In order to keep $\epsilon$ low, we must implement key renewal strategy in Keystone. In fact, OpenStack has this strategy implemented and by default Keystone's key is changed every one hour. 

\textit{Statement.}
\textit{The implementation of the discussed and proven secure $RAF$ mechanism provided in section \ref{sec:implementation}, is a one-time token}

In order to have a one-time $RAF$ token, we have considered and implemented an additional component; Blacklist. The blacklist, ensures that each module cannot request a validation for the same token more than one time.%the token is not been used by the requested service more than one time. 
The blacklist component must be added to the authentication module and functions as follows: Each time the authentication module receives a token, first, it checks if the token is valid, i.e. the $Ver$ function outputs $1$, if so, then it checks if the pair (base-RAF, service id) is in the blacklist. If not, then the authentication module returns valid, otherwise returns invalid (for more detail on blacklist functionality see section \ref{sec:backlist}).

This indicates that each token with the same base-RAF token can only be validated once, which implies that the RAF token implementation is a one-time token.

%Since our implemented $RAF$ token in section \ref{sec:implementation} is a one-time token, then it fits the security proof of OpenStack given in \cite{hogan2019} and we can adopt the proven security. 
%Hogan et al. \cite{hogan2019} provided a universally composable security analysis of OpenStack. The authors proved that if instead of bearer token a one-time token is used, the security properties of OpenStack will improve (for more details on OpenStack security analysis, see \cite{hogan2019}). 

%Therefore, using our RAF mechanism instead of bearer token will improve the security of OpenStack. Note that the proof in \cite{hogan2019} is done via universal composability framework. This framework guarantees that a component remains secure even if arbitrarily composed with other component. This means that using the RAF improves the security of OpenStack and won't be affected if in the future new modules are added to the system. 

\subsection{The Security Analysis of Fully Tied RAF}\label{sec:AnalysisFullyRAFT}
The fully tied RAF is similar to user tied RAF except it has an additional feature which prevents the adversary to send a token on behalf of a service. In other words, if the adversary is able to get access to a fully tied token, s/he is not able to send a request on behalf of a (uncorrupted) service to another service.

In this section we show the security guarantees of a fully tied RAF. We first provide a technical definition of a fully tied RAF token. Next, we define a game between the adversary and the fully tied RAF: In this game the adversary is given an oracle access to the RAF function and is able to make queries. At some point, the adversary generates a token for a message that he has not queried. The adversary wins the game if the provided token passes the validation.\\
We also give the security definition of fully tied RAF and construct a secure fully tied RAF using the algorithms in section \ref{sec:generating}. Finally, we wrap up this section with a lemma that shows our construction is secure against chosen-message attack.

\begin{definition}\label{def:fully-tiedToken}

A Fully-tied Token is a quadruple $\Pi= (Gen, UserIssue, ServiceIssue, Ver)$ of ppt algorithms where:
\begin{itemize}
    \item $K=(k_0,\cdots,k_{n-1})\leftarrow Gen(1^\lambda)$ with security parameter $\lambda$. %\textcolor{red}{K should not be represented as a set, it is a tuple.} %We denote $K=\{k_1,\cdots,k_{n-1}\}$
    \item $r_0=(m_0,t_0) \leftarrow UserIssue(k_0,p,c_0)$ generates a token $r_0$ for root payload $p$ and the command $c_0$ with key $k_0$. 
    \item $r_j=(m_j,t_j) \leftarrow ServiceIssue(i, K,r_{j-1},c_j)$  internally call $Ver(K,r_{j-1})$ and if $r_{j-1}$ is invalid then it returns nothing; else
    generates a token $r_j$ for the input token $r_{j-1}$, the service id $i$, 
    and the command $c_j$ with the key  $k_i$. Note that  $ServiceIssue$ consists of two phases: 1- Verification: in which it calls $Ver$ algorithm and verifies the validity of the token $r_{j-1}$. For this part $K$ is needed. 2- Token generation: computes a token $r_j$ using the key of the service with service id $i$ (i.e. $k_i$).

     \item $b \leftarrow Ver(K,r_n=(m_n,t_n))$, where $Ver$ takes a set of keys $K$ and a token $r_n$. It outputs $b$, with $b=1$ meaning the token is valid and $b=0$ meaning the token is not valid.
     \item It is called correct if for every $K$ output by $Gen(1^\lambda)$ and every token $r$ generated by $UserIssue$ or $ServiceIssue$; it holds $Ver(K,r)=1$.
\end{itemize}
\end{definition}
According to the above definition, a fully tied token gets a payload and extends it with a command. This definition looks similar to definition \ref{def:user-tiedToken}. However, an important difference exists: here, each module has its own secret key, while in definition \ref{def:user-tiedToken}, modules do not own/use any secret keys. This enable modules to add a new command to the token and sign the token with their secret keys.
%Also, each module has a secret key. Modules can also add a new command to the token and signs the token with its secret key. 

We define the following game for a PPT adversary ${\cal A}$, the security parameter $\lambda$, and a fully-tied token $\Pi = (Gen, UserIssue, ServiceIssue, Ver)$:

\begin{game} \textbf{Forge a Fully Tied Token or for short ForgeFTT}:\label{gm:FTT}
\begin{itemize}
    \item $K=(k_0,\cdots,k_{n-1})\leftarrow Gen(1^\lambda)$ %\textcolor{red}{same comment as above} 
    where $n=$Number of Services in the system.
    \item The adversary $\cal A$ is given input $1^\lambda$ and access to oracles: $Ver(K,.)$, $UserIssue(k_0,.,.)$, and $ServiceIssue(i,K,.,.)$ and knows keys $k_i$ for $i \in I\subseteq \{1,\ldots, n\}$.  
    Let $\Phi$ represent the set of all tokens that $\cal A$ has requested. Eventually, the adversary outputs $r=(m,t)$.
    \item $\cal A$ wins if and only if (1) $Ver(K,r=(m,t))=1$, (2) $r \notin \Phi$, and (3) $r$ is not an output of $ServiceIssue(i,K,.,.)$  for some $i\in I$.
    \item The experiment returns 1 if the adversary wins the game, otherwise it returns 0.
\end{itemize}
\end{game}

This game is exactly the same as game \ref{gm:UTT} with one extra condition in Step 3, which is "$r$ is not an output of $ServiceIssue(i,K,.,.)$  for some $i\in I$".

\begin{definition}
A fully tied token $\Pi = (Gen, UserIssue, ServiceIssue, Ver)$ is unforgeable under a chosen-message attack, or just secure, if for all PPT adversaries $\cal A$
there exists a negligible function \textbf{negl} such that:
\begin{equation} 
    \Pr[ForgeFTT_{{\cal A}, \Pi} (\lambda)=1]\leq negl(\lambda)
\end{equation}

\end{definition}
A secure fully-tied token guarantees the authenticity of the user and service commands embedded in a token, i.e., not only the adversary can not forge a token with a fake user-command, he also can not forge a valid token on behalf of any service, unless he  corrupts the service (knows the secret key of the service).

\begin{lemma}\label{lemma:1}
Assume that the HMAC used in Construction \ref{con:FulRAFT} is secure and indistinguishable from a random function.  
Then, Construction \ref{con:FulRAFT} is a secure fully tied token that is unforgable under chosen message attacks. 
\label{lem:FullRaft}
\end{lemma}

\begin{construction}\textbf{Fully Tied Token RAF} \label{con:FulRAFT}

Let HMAC be indistinguishable from a random number. Define a Fully Tied Token $\widehat{RAFT}=(Gen, UserIssue, ServiceIssue, Ver)$ as follows:
\begin{itemize}
    \item \textbf{$Gen(1^\lambda)$}: Upon input $1^\lambda$, choose $K=(k_0, k_1, ..., k_{n-1}) \leftarrow \{0,1\}^{\lambda}$  %\textcolor{red}{Here and below represent K as a tuple.} $K=\{k_0, k_1, ..., k_{n-1}\} \leftarrow \{0,1\}^{\lambda}$ 
    where $n=$Number of Services in the system. 
    \item \textbf{$UserIssue(k_0,p,c)$}: Upon input key $k_0 \in \{0,1\}^{\lambda}$, payload $p$ and the command $c$ compute $r$ by running Algorithm \ref{alg:GenRAFTUser}.
    \item \textbf{$ServiceIssue(i,K,r',c)$}: Upon input key  %$k_i \leftarrow \{0,1\}^{\lambda}$ 
    $K=\{k_0, k_1, ..., k_{n-1}\} \leftarrow \{0,1\}^{\lambda}$
    , token $r'$ and the command $c$ compute $r$ by running Algorithm \ref{alg:GenFRAFTService}. %\textcolor{red}{The input is back to $i$ and $K$.}
    \item \textbf{$Ver(K,r=(m,t))$}: Upon input key set $K=\{k_0, k_1, ..., k_{n-1}\} \leftarrow \{0,1\}^{\lambda}$, token $r$ compute $b$ by running Algorithm \ref{alg:verRAFTFull}.
\end{itemize}
\end{construction}
%%%%%%%%%%%%%%%%%%%%%%%%%%%%%%%%%%%%%%%%%%
%%%%%%%%%%%%%%%  BEGIN  %%%%%%%%%%%%%%%%%%
%%%%%%%%%%%%%%%%%%%%%%%%%%%%%%%%%%%%%%%%%%

We defer to Appendix \ref{app:proof} for the full proof. Here, we simply highlight the main ideas involved in the proof.

\begin{proof}[Proof sketch]
The intuition behind the proof of this lemma is the same as lemma \ref{lem:UTRaft1}. Here, our argument also proceeds via reduction approach. 

We argue that the only way for an adversary to win the game is to either generate a valid token which is a result of $UserIssue$ or a valid token which is the result of $ServiceIssue$. Next, we show that a tokens generated by $UserIssue$ can be represented as a user tied RAF token with exactly one command and tokens generated by $serviceIssue$ can be represented by a user tied RAF token with no command. This argument enables us to use lemma \ref{lem:UTRaft1} to prove the security of the fully tied RAF and show its upper bound.
\end{proof}

\textit{Discussion.} For fully tied RAF the upper bound on the probability of successfully forging a token is much more tight than the upper bound for  user tied RAF. This is reasonable because in this type of RAF mechanism, each service has its own secret key for applying the $HMAC$. This means even if the adversary gets access to a valid token, he is not able to generate a valid token on behalf of an uncorupted service.

%% file: sections/implementation.tex
\section{Proof of Concept}\label{sec:implementation}
To illustrate a proof of concept, we utilize OpenStack as the testbed for our proposal. We have developed a Python library that enables us to:
\begin{itemize}
    \item Issue a user-tied RAF token, given a Fernet token or another user-tied RAF token.
    \item Validate a user-tied RAF by providing the Fernet key and retrieving all commands within the user-tied RAF token.
\end{itemize}

We also modified Keystone to accept and validate user tied RAF tokens. In the rest of this chapter, we explain the details of our experiments. 

\subsection{Experiment platform} \label{sec:platform}
Using KVM, we set up a Virtual Machine (VM) with 4 vCPUs and 8GB vRAM.  The physical host was a DELL XPS 9530 laptop taking advantage of an Intel Core(TM) i7-4712HQ CPU @ 2.30GHz with 16GB memory and 250GB SSD. The operating system of the host and virtual machine were Ubuntu 18.04.3 LTS. We had installed OpenStack Stein on our VM using DevStack script.

\subsection{RAFT Library}
Snippet \ref{code:RAFTLib} represents the user tied RAF library. It contains two classes: 
\begin{itemize}
    \item KeystoneRaft: this class encapsulates several methods to validate RAF tokens. Some of its important methods are:
    \begin{itemize}
        \item isRAFT: gets a token and returns true if the token was a RAF token.
	    \item CheckExpirationTime: gets a token and checks its expiration time. If the token is expired, then it raises an exception error.
	    \item GetKey: This method is used internally to find the key of the token. 
	    \item ValidateRAFT: gets a RAF token and checks if it is valid or not. If the token was valid, this function returns all the commands inside the token. Also, it sets the fernet\_token property of the class.
	\end{itemize}
	\item ClientRaft: this class allows clients and other modules to create a new RAF token for their requests. The main methods of this class are:
	\begin{itemize}
	    \item \_\_init\_\_: this is the constructor of the class and allows us to define the parent token of the RAF token, which we want to create.
        \item SetParentToken: Allows to set or change the parent token.
        \item SetCommand: Is used to set the command that will be a part of the token. 
        \item Finalize: generates and returns a RAF token based on the parent token and the command that had been set. 
	\end{itemize}
\end{itemize}

\subsection{The required changes}
If you install OpenStack with its default setting, the source code of Keystone will be at \mycmd{/opt/stack/keystone} folder. Let's call this folder the \textit{base folder}. Then, our library must be added to the following folder:
\begin{mycode}
base folder/keystone/token
\end{mycode}
The source code of Fernet token must be at:
\begin{mycode}
base folder/keystone/token/providers/fernet
\end{mycode}
In the \mycmd{fernet} folder, there is a file \mycmd{token\_formmater.py}, which contains the entry point for the Fernet library. We added the following code at the library importing section of the file (at the beginning of the file):
\begin{mycode}
import RAFT
\end{mycode}
Then, we modified the \mycmd{validate\_token} function of \mycmd{TokenFormatter} class, as shown in snippet \ref{code:validateToken}. When this function is called, we first check the input token. If it is a RAF token, then we process it using an instance of \mycmd{KeystoneRAFT} class. Upon successful validation, we extract the original Fernet token and assign it to the token. In this way, the rest of the function checks the privileges of the root Fernet token. At the end of the function, if the input token was a RAF token, we adjust the expiration time of token according to the shortest expiration time in the RAF token.  

After applying the changes to source code, the Keystone service must be restarted. Do not restart the virtual machine. If you do so, you need to uninstall OpenStack and try to install it again. In order to restart the Keystone service execute the following command in the Linux terminal. 
\begin{mycode}
systemctl restart devstack@keystone.service
\end{mycode}
\subsection{Comparison Between RAF and Fernet Tokens' Creation and Validation Time }
Table \ref{tab:firstSetRF} shows the average time (in milliseconds) needed for creation and verification of a token, deviation, deviation to above, and deviation to bellow the average. If we use the Fernet token mechanism, the creation and validation time is always near the average in row 1. In the RAFT mechanism, whenever a user wants to issue a new command, he needs to create a new RAF token from a Fernet token. By default, each Fernet token is valid for an hour. Hence, when the user grants a Fernet token, he can use it to create new RAF tokens locally for one hour. In the statistic about RAF, we also include the time of granting a Fernet token.

\newcolumntype{L}{>{\arraybackslash\columncolor[HTML]{F5F5FF}}m{4.3cm}}
\newcolumntype{t}{>{\columncolor[HTML]{F5F5FF}} l}
\newcolumntype{f}{>{\columncolor[HTML]{EEFFEE}} c}
\begin{table*}[ht]
    \centering
%    \rowcolors{2}{red!5}{red!2}
    \begin{tabular}{|c|L|c c c c |f|}
        \hline
\rowcolor{lightgray} \# & Token Type & Average & dev. & dev. to above & dev. to bellow  \\
\hline
1 & Fernet &	76.54 & 9.39 & 8 & 0 \\
2 & RAF, 1\textsuperscript{\P} & 79.24 & 13.38 & 4 & 2 \\ %& 103.5
3 & RAF, 2 & 56.79 & 3.02 & 12 & 11 \\ %& 74.2
4 & RAF, 10 & 40.83 & 1.57 & 9 & 6 \\ % & 53.3
5 & RAF, 50 & 37.88 & 0.81 & 13 & 15 \\ %& 49.5
6 & RAF, 100 & 37.21 & 0.64 & 13 & 13 \\ %& 48.6
\hline
\rowcolor{white} \multicolumn{6}{l}{\textsuperscript{\P}: The number indicates the number of tokens issued per hour.}\\
    \end{tabular}
    \caption{Execution time in milliseconds for generating and validating of a token based on 100 sample.}
    \label{tab:firstSetRF}
\end{table*}

As mentioned earlier, one idea for mitigating the bearer token problem is to shorten the lifetime of Fernet tokens in a way that there is no time to use it twice. As our experiment shows, if the user issues more than one command in an hour (rows 3 to 6), the RAF mechanism works 30 to 50 percent better than the idea of shortening the lifetime of Fernet tokens.

Table \ref{tab:firstSetCL} shows the length of the command that we add to a RAF token does not noticeably effect the average time (in milliseconds) of creation and verification of a token. For this experiment, we create RAF tokens for hypothetical commands with 1, 100, 200, and 1000 character length. 

\begin{table*}[ht]
    \centering
%    \rowcolors{2}{red!5}{red!2}
    \begin{tabular}{|c|L|c c c c |f|}
        \hline
\rowcolor{lightgray} \# & Command length & Average & dev. & dev. to above & dev. to bellow  \\
\hline
1 & 0 & 40.11 & 5.89 & 8 & 1 \\
2 & 100 & 41.01 & 9.7 & 6 & 0 \\
3 & 200 & 39.34 & 3.94 & 14 & 7 \\
4 & 1000 & 40.78 & 5.02 & 12 & 8 \\
\hline
    \end{tabular}
    \caption{The effect of the length of the command (in character) that is added to a RAF token in generation and verification time in milliseconds based on 100 sample.}
    \label{tab:firstSetCL}
\end{table*}

\subsection{Comparison Between RAF and Fernet Tokens' Creation and Validation Time }

The RAF solution puts a negligible processing overhead over the  Fernet processing time. Since many processes were running on the OpenStack server, this negligible overhead was not measurable using OpenStack. Hence, we calculated the overhead locally. In our platform \ref{sec:platform}, issuing 100 RAF tokens (excluding granting a Fernet token from Keystone) took 8.1 milliseconds, and verification of just the RAF tokens (excluding the verification of the base Fernet) took 13.2 milliseconds. These numbers show the extra processing time needed for a RAF token compare to Fernet token.

Table \ref{tab:firstSetExp} shows the summary of the first set of our experiments (see the source code in Snippet \ref{code:firstExp}) that focuses on the creation and validation time of user tied RAF and Fernet token \footnote{The process of issuing and validating a fully tied RAF token with one command and a user tied RAF token with one command is exactly the same. 
The only difference between issuing and validating a fully tied RAF and a user tied RAF token for more than one commands is that they use different keys for the HMAC of second and more commands. Hence, we do not expect any considerable change in the result if we repeat them for fully tied RAF.}. The first set includes  the following seven experiments:
\begin{itemize}
    \item Row \#1 shows the execution time for getting 100 Fernet tokens from Keystone.
    \item Row \#2 represent the execution time for getting a Fernet token and generating 100 RAF tokens using the Fernet token.
    \item Row \#3 exposes the execution time for getting and validating of 100 Fernet tokens.
    \item Rows \#4 to \#7 demonstrate the execution time of getting a Fernet token and issuing 100 RAF tokens with hypothetical commands with 0, 30, 60, and 200 lengths (in character). 
\end{itemize}
To get more reliable results, we repeat each experiment several times, but because space limitation, we only represent the result of the experiments. 

\begin{table*}[ht]
    \centering
%    \rowcolors{2}{red!5}{red!2}
    \begin{tabular}{|c|L|c c c c c |f|}
        \hline
\rowcolor{lightgray} \# & Experiment title & First & Second & Third & Forth & Fifth & Avg \\
\hline
1 & Fernet Generation & 4.3519 & 4.3519 & 4.3593 & 4.4529 & 4.4897 & 4.4011 \\
2 & RAFT Generation & 0.0489 & 0.0519 & 0.0510 & 0.0505 & 0.0481 & 0.0501 \\
3 & Fernet Generation and validation & 8.9275 & 8.8489 & 8.8471 & 8.5349 & 8.8222 & 8.7961 \\
4 & RAF token Generation and validation; 0L & 4.6791 & 4.5629 & 4.6725 & 4.6074 & 4.6696 & 4.6383 \\
5 & RAF token Generation and validation; 30L & 4.6528 & 4.6620 & 4.5986 & 4.6332 & 4.6296 & 4.6353 \\
6 & RAF token Generation and validation; 60L & 4.6668 & 4.5719 & 4.6438 & 4.7847 & 4.5213 & 4.6377 \\
7 & RAF token Generation and validation; 200L & 4.6302 & 4.6704 & 4.7203 & 4.5572 & 4.6420 & 4.6440 \\
\hline
    \end{tabular}
    \caption{Execution time in seconds for generating and validating of 100 RAF tokens  and Fernet tokens. For creating RAF tokens, we only interact with Keystone one time and then create 100 RAF tokens locally.}
    \label{tab:firstSetExp}
\end{table*}

As mentioned earlier, one idea for mitigating the bearer token problem is to shorten the lifetime of Fernet tokens in a way that there is no time to use it twice. From row \#1 and \#2,  we can conclude that for token generation, if a user issues more than one command in a specific period, the RAF solution is faster than the short-life Fernet token idea. But the creation time is only one side of the problem. Each token needs to be validate by Keystone. If we include the validation time (rows \#3 to \#7), the RAF solution still works at most 50 percent better if each user issues several commands.

The length of the command that we add to a RAF token has unnoticeable overhead, as understood from rows \#4 to \#7. In fact, we repeated the experiment \#4 to \#7 much more than five times, and each time we got very close results. 

\subsection{RAF and Fernet in Action}
In the second set of experiments, we examined the effect of using user tied RAF instead of Fernet tokens in the execution time of four different commands as follows:

\noindent \textbf{Create volume}: is a moderate workload command that creates a volume, i.e., a permanent memory like a physical disk. Snippet \ref{code:createVol} shows the sample command that we used for creating a volume. We did not load any image to the volume. If we want to load an image to the volume, it can be considered as a high load task. 

\noindent \textbf{Create VM (server)}: is a typical command in OpenStack that consumes significant resources. Snippet \ref{code:createVM} shows the sample command that we used for creating a VM. In this sample, we used \mycmd{cirros} that is an ultra-lightweight operating system. The \mycmd{cirros} operating system (OS) includes a few core functionalities of a Linux OS and only be used for tests.  

\noindent \textbf{Image list}: is a light load command that does not need any parameter. 

\noindent \textbf{Project list}: is another light load command that returns the list of projects that are available for the owner of the token.  This command also does not need any parameters.

Table \ref{tab:SecondSetExp} represents a comparison between the execution time (in seconds) of the four different commands using RAF and Fernet tokens. We put the result of five execution for each command in the table. Based on the result, the overhead of using RAF instead of Fernet in typical create-server and create-volume commands is less than 1 percent. Here, we used cirros OS to create a server that is a very small nonpractical operating system. In a real deployment, usually creating a VM using an operational OS takes several seconds, and the RAF overhead is not noticeable in such cases. In general, typical cloud operations like creating servers, migrating servers, and taking back up of servers are very time consuming compare to token verification. Hence, we expect that this overhead will be much less than 1 percent for a real deployment. 

%\textcolor{red}{The table clearly does not include a Fernet token generation, right. In the RAFT case, RAFT generation by the service is included? It is not exactly clear to me which computations are included in the execution times.}

\newcolumntype{Z}{>{\arraybackslash\columncolor[HTML]{F5F5FF}}m{3.5cm}}
\begin{table*}[ht]
    \centering
%    \rowcolors{2}{red!5}{red!2}
    \begin{tabular}{|Z|c c c c c |f|f|}
        \hline
\rowcolor{lightgray}  Experiment title & First & Second & Third & Forth & Fifth & Avg & Ratio\\
\hline
Creating a volume (Fernet) & 0.3243 & 0.3681 & 0.4546 & 0.4396 & 0.4246 & 0.4022 & \multirow{2}{*}{1.0078} \\
Creating a volume (RAF) & 0.3583 & 0.4021 & 0.4140 & 0.4313 & 0.4211 & 0.4054 & \\
\hline
Creating a VM (Fernet) & 0.5208 & 0.5966 & 0.8908 & 2.3995 & 0.6762 & 1.0168 & \multirow{2}{*}{1.0030} \\
Creating a VM (RAF) & 0.7032 & 0.7740 & 0.8938 & 2.0452 & 0.6831 & 1.0198 &\\
\hline
Getting image list (Fernet) & 0.0584 & 0.0571 & 0.0569 & 0.0605 & 0.0526 & 0.0571 & \multirow{2}{*}{1.0280} \\
Getting image list (RAF) & 0.0548 & 0.0598 & 0.0616 & 0.0553 & 0.0620 & 0.0595 & \\
\hline
Getting project list (Fernet) & 0.0293 & 0.0332 & 0.0274 & 0.0296 & 0.0294 & 0.0298 & \multirow{2}{*}{1.0326} \\
Getting project list (RAF) & 0.0303 & 0.0323 & 0.0296 & 0.0311 & 0.0305 & 0.0307 & \\
\hline
    \end{tabular}
    \caption{Comparison between the execution time of four different commands using RAFT and Fernet tokens}
    \label{tab:SecondSetExp}
\end{table*}

Snippet \ref{code:secExp} shows the source code that we developed for the second set of experiments.

%% file: sections/summary.tex
\section{Summary} \label{sec:summary}
Modularity and modular thinking in developing extensive systems have a pivotal role. Robust authentication and authorization mechanisms within such modular frameworks are of importance. This paper focuses on token-based authentication mechanisms, where users present credentials to obtain tokens for subsequent interactions with other modules. Tokens have advantages over traditional username and password authentication, focusing on enhanced security through reduced password exposure, token expiration, and scoped access. However, they suffer from a critical vulnerability known as the "bearer token" problem, where the compromise of user tokens poses significant security risks in modular systems, leading to a trade-off between system performance and security.

To address the bearer token problem, the paper introduces a novel token-based authentication system RAF, which complies with the Modularity Security Requirement, advocating for a scenario in which, if an adversary corrupts one module, other modules should only execute user-requested actions without additional privileges. RAF includes RAF tokens, a Policy Enforcer (PE) component, and a Blacklist component. The main idea involves users receiving Fernet tokens from the identity module, generating a RAF token for every command, embedding the command in the token, and using it instead of Fernet tokens. The identity module validates RAF tokens in a recursive manner and extracts all the commands embedded in the token. The PE ensures command consistency, and the Blacklist prevents replay attacks. RAF tokens have a determined structure, including commands, and are single-use tokens. Hence, an RAF token connects the token and the reason for creating the token. 

The paper proposes two variations of RAF: User-tied RAF and Fully-tied RAF, each with distinct security features and requirements. The verification process and token generation algorithms are detailed. The Blacklist prevents replay attacks, while the Policy Enforcer ensures command consistency, completing the solution. The paper discusses the advantages and challenges of User-tied RAF and Fully-tied RAF, emphasizing security and scalability considerations. 

User-tied RAF tokens are more straightforward to implement and do not require services to have secret keys. Each module can independently issue RAF tokens without dependencies on service-specific keys. The absence of service-specific keys reduces the complexity of key management. They can be quickly adopted in existing systems without significant changes, making them backward compatible with systems using traditional Fernet tokens. On the other hand, since they do not tie tokens to specific services, an adversary who obtains a token can potentially forge child tokens. However, the Policy Enforcer helps mitigate this by ensuring command consistency.

Fully-tied RAF tokens provide a higher level of security by tying each token to a specific service using a service-specific secret key. Fully-tied RAF ensures that tokens are valid only for the intended service and command. Hence, it provides additional protection against module corruption. An adversary cannot forge a valid RAF token on behalf of a module unless the module itself is compromised. On the other hand, Fully-tied RAF introduces the challenge of distributing and managing secret keys for each service. The need for each module to have a secret key shared with the identity module may pose scalability challenges, particularly in large and dynamic environments. Consequently, adopting Fully-tied RAF may require significant changes in existing systems due to the need for service-specific secret keys. 

The paper discusses the security guarantees of the two variations of RAF. It outlines these tokens' technical definitions, security analyses, and construction. The security analysis relies on two game-based proofs, assuming HMAC is a pseudo-random function. Using the games, the paper shows that if adversaries can break RAF tokens with any probability more than negligible, they can break the security of HMAC.

The paper demonstrates a proof of concept for User-tied RAF tokens within the context of OpenStack. The implementation involves modifications to Keystone, the creation of an RAFT library, and provides details about the experimental platform and results.

The results include a comparison of the creation and validation times of RAF and Fernet tokens under various scenarios, as well as the execution time of specific OpenStack commands using RAF and Fernet tokens. The findings indicate minimal overhead in typical scenarios, showcasing the viability of RAF tokens. Additionally, the results highlight the significant advantage of using RAF over short-life Fernet tokens.

%% file: app/AppProof.tex
\section{RAFT Security Guarantees}\label{app:proof}
\subsection{Proof of Lemma \ref{lem:UTRaft1}}
%%%%%%
Based on Game \ref{gm:UTT} we define two other games which we will need for the proof of Lemma \ref{lem:UTRaft1}:
\begin{itemize}
    \item Consider $ForgeUTT^i_{{\cal A},\Pi}$ to be the notion for forging a token with exactly $i$ commands. For this, we can define a game that is exactly as game \ref{gm:UTT} with one extra condition in Step 3, which is \textit{"m contains exactly $i$ commands"}
    \item Consider $ForgeUTT^{\leq i}_{{\cal A},\Pi}$ to be the notion for forging a token with at most $i$ commands. For this, we can define a game that is exactly as game \ref{gm:UTT} with one extra condition in Step 3, which is \textit{"m contains at most $i$ commands"}
\end{itemize}

\begin{proof}
The intuition behind the proof of this lemma is that for a PPT adversary, forging a user tied RAFT without observing any of its ancestors involves distinguishing the HMAC from a random function.%\textcolor{green}{will mention at the begining of this chapter and be consistant through the whole chapter.} %\textcolor{red}{did you put the removed sentence somewhere else where you use this notation for the first time? }%To make the proof easier to follow, we write $HMAC(k,.)$ instead of $HMAC_k(.)$;  the first parameter $k$ is the key of the HMAC. \textcolor{red}{you have already used this notation in the lemma, so you need to write this before the lemma or where you used this notation the first time} 
We simplify the descriptions of algorithms \ref{alg:GenRAFTUser}, \ref{alg:GenURAFTService}, and \ref{alg:verRAFTUser} as follows:
\begin{itemize}
    \item For the key $k$ and input payload $p$ and the command $c$\\
    $UserIssue(k,p,c) = HMAC(HMAC(k,p), p\parallel c)$
    \item For the key $k$, input token $r=(m,t)$, and the command $c$
          \[ ServiceIssue(k,r=(m,t), c) =
            \begin{cases}
            HMAC(t, m\parallel c)       & \quad \text{if } Ver(k,r) \text{ is valid}\\
            nothing  & \quad \text{if } Ver(k,r) \text{ is  invalid}
            \end{cases}
        \]
    \item For the key $k$, and the input token $r=(m,t)$, first, $Ver(k,r)$ decomposes $m$ and gets $\{p,c_1,\cdots,c_n\}$, then calculates: 
        \begin{align*}
            h_0 &= HMAC(k,p)\\
            h_1 &= HMAC(h_0, p \parallel c_1) \\
            \vdots\\
            h_n &= HMAC(h_{n-1}, p \parallel c_1 \parallel \cdots \parallel c_n)
        \end{align*}
    \noindent Then, if $h_n = t$, returns \textit{valid}; otherwise, returns \textit{invalid}.
\end{itemize}
%\textcolor{green}{AGREED -- after the next two paragraphs, do make sure the text flows. == In my opinion we need to move this paragraph to after we define RAFT'}

%\textcolor{red}{We also introduce the notation $RAFT(k,p,c_1,\ldots,c_i)$ respectively $RAFT'(k,p,c_1,\ldots, c_i)$ to stand for ... make this general and define $RAFT(k,p)=h_0=HMAC(k,p)$ etc. }

Let ${\cal A}$ be a probabilistic polynomial-time adversary and let $\varepsilon ($.$)$ be a function
such that:
\begin{equation} \label{eq:3}
\Pr[ForgeUTT_{{\cal A}, RAFT} (\lambda)=1]= \varepsilon (\lambda)
\end{equation} 
and let $\Phi$ represent the set of all tokens that $\cal A$ has queried.
Now, we define a new scheme $RAFT'=(UserIssue', ServiceIssue', Ver')$, which is similar to $RAFT$ except a random function $f(.)$ is replaced with the first HMACs in $UserIssue$ and $Ver$ algorithms. In other words, \textit{HMAC(k,p) is replaced by f(p)}. %\textcolor{red}{What is in brackets is actually the important part and should be upfront.}

We show that this implies the existence of a polynomial-time algorithm that can distinguish the HMAC from a random one with advantage $O(\varepsilon)$. This will then imply that  $\varepsilon$ must be negligible, as required.

We define the notation $RAFT(k,p,c_1,\ldots,c_i)$ respectively $RAFT'(k,p,c_1,\ldots, c_i)$ to stand for a RAFT/RAFT' token with $i$ commands and define $RAFT$ and $RAFT'$ as follows:
\[
RAFT(k,p,c_1,\ldots, c_i) = 
\begin{cases}
  \resizebox{.60\hsize}{!}{$HMAC(RAFT(k,p,c_1,\ldots,c_{i-1}),p\parallel c_1\parallel \ldots \parallel c_i)=h_i$} & i>0 \\
  HMAC(k,p)=h_0 & i= 0
\end{cases}
\] \label{eq:RAFTEq}
and
\[
RAFT'(k,p,c_1,\ldots, c_i) = 
\begin{cases}
  \resizebox{.60\hsize}{!}{$HMAC(RAFT'(k,p,c_1,\ldots,c_{i-1}),p\parallel c_1\parallel \ldots \parallel c_i)=h_i$} & i>0 \\
  f(p)=h_0 & i= 0
\end{cases}
\]

From $\cal A$ we will now construct a polynomial-time distinguisher $\cal D$ that is given an oracle $\cal O$ that is either of $HMAC(k,.)$ or a random function $f(.)$. The construction $\cal D$ works as follows:

\begin{itemize}
    \item Upon input $\lambda$, adversary $\cal D$ passes $\lambda$ to $\cal A$.
    \item When $\cal A$ queries its oracle $UserIssue(k,.,.)$ with a message $m'=(p',c')$, $\cal D$ queries $\cal O$ with $p'$ and receives the result $h'$. Next, it calculates the value $t'=HMAC(h',m')$ and hands it to $\cal A$ and continues.
    \item When $\cal A$ queries its oracle $Ver(k,.)$ with a message $r''=(m'',t'')$,
    $\cal D$ runs algorithm \ref{alg:verRAFT}. Algorithm \ref{alg:verRAFT} is exactly the same as algorithm \ref{alg:verRAFTUser} except that instead of using HMAC to compute $h_0$, it uses the oracle $\cal O$. Upon receiving $b''$ from algorithm \ref{alg:verRAFT}, $\cal D$ hands it to $\cal A$. %\textcolor{red}{same comment -- you have an algorithm 7prime} %and receives $b''$. Then, $\cal D$ hands $b''$ to $\cal A$. 
    \item When $\cal A$ queries its oracle $ServiceIssue(k,.,.)$ with a message $(r'=(m',t'),c')$, $\cal D$ runs algorithm \ref{alg:verRAFT} with $r'$ and gets the response $b'$. 
    %\textcolor{red}{explain what algorithm 7 is doing (swap the next bullet with this bullet) -- actually, it is not exactly algorithm 7 since it asks the oracle to produce $h0$ (it seems you may want to explain that algorithm 7 with access to the oracle is reflected by the pseudocode of algorithm 9) -- when you explain this, you can refer back to the chain description which will make it easier for the reader to understand} 
    If $b'=0$, $\cal D$ does nothing. However, if $b'=1$, $\cal D$ hands $HMAC(t',m'\parallel c')$ to $\cal A$ and continues.
\end{itemize}
At the end, when  $\cal A$ outputs $r=(m,t)$,  $\cal D$ runs algorithm \ref{alg:verRAFT} %\textcolor{red}{see above} 
and receives $b$. If $b=1$ and  $r$ and any of its ancestors are not in $\Phi$%$r$ and none of its ancestors asked before by $\cal A$ 
%\textcolor{red}{make this precise by keeping a list $\Phi$ (as in game 1) in the above abstract description of $\cal A$}
, then $\cal D$ outputs 1. Otherwise it outputs 0.

\begin{algorithm}
\SetKwProg{Fn}{Function}{}{}
\Fn{Verify(r)}
{
 $(p,c_0, \cdots, c_n, t) \leftarrow Unpack(r)$ \\
 $i=0$ \\
 $h_0= {\cal O}(p)$\\
  \While{$i\leq n$}{
  $h_i=HMAC(h_{i-1}, p \parallel c_0 \parallel \cdots \parallel c_i)$\\
  $i=i+1$
  }
 \eIf{$h_n = t$}
 {
     return 1;
 }
 {
    return 0;
 }
 }
 \bf{End Function}\\
  \caption{Pseudocode of verifying}
  \label{alg:verRAFT}
\end{algorithm}
Since $\cal A$ runs in polynomial time, $D$ also runs in polynomial time. From the construction of $\cal D$, it is clear that depending on $\cal O$,  $\cal A$ either plays $ForgeUTT_{{\cal A},RAFT}$ or $ForgeUTT_{{\cal A},RAFT'}$, and we have:
\begin{align}
    \Pr[ForgeUTT_{{\cal A},RAFT}(\lambda)=1] &= Pr[D^{HMAC(k,.)}(\lambda)=1]\label{eq:raftDis}\\
    \Pr[ForgeUTT_{{\cal A},RAFT'}(\lambda)=1] &= Pr[D^{f(.)}(\lambda)=1]\label{eq:raftpDis}
\end{align}
From \ref{eq:distinguish1}, \ref{eq:raftDis}, and \ref{eq:raftpDis}:
\begin{align}
    |\Pr[ForgeUTT_{{\cal A},RAFT}(\lambda)=1] - \Pr[ForgeUTT_{{\cal A},RAFT'}(\lambda)=1]| \leq \alpha(\lambda) \label{eq:raft_raftp}
\end{align}

%Consider $ForgeUTT^i_{{\cal A},\Pi}$ to be the notion for forging a token with exactly $i$ commands. For this, We can define a game that is exactly as game \ref{gm:UTT} with one extra condition in Step 3, which is \textit{"m contains exactly $i$ commands"}. 
The analysis presented so far can be repeated in exactly the same way for the game $ForgeUTT^i$. This gives:
%Then we can say:
%a game which indicates forging a token with exactly $i$ commands. This game is exactly the same as game \ref{gm:UTT} except with the following additional constraint in the third step, \textit{(3) m be exactly $i$ commands.}
 %\textcolor{red}{Also refer to how where the game is changed -- in what line an extra condition is added. Explain that if you change the above algorithm $D$ by also checking on whether $r$ has exactly $i$ commands, then all the arguments remain the same and you get the next inequality.} So,
%Consider $ForgeUTT^i_{{\cal A},\Pi}$ means forging a token with exactly $i$ commands. \textcolor{red}{Also refer to how where the game is changed -- in what line an extra condition is added. Explain that if you change the above algorithm $D$ by also checking on whether $r$ has exactly $i$ commands, then all the arguments remain the same and you get the next inequality.} So,
\begin{align}
    |\Pr[ForgeUTT^i_{{\cal A},RAFT}(\lambda)=1] - \Pr[ForgeUTT^i_{{\cal A},RAFT'}(\lambda)=1]| \leq \alpha(\lambda) \label{eq:raft_raftpi}
\end{align}
%%\noindent Lets consider the following game:
%%\begin{game} \textbf{Forge an HMAC $Forge_{\cal A'}HMAC (\lambda)$:}\label{gm:HMAC}
%%\begin{enumerate}
%%    \item $Gen(1^\lambda)$ generates a key $k$
%%    \item The adversary $\cal A'$ is given input $1^\lambda$ and oracle access to $HMAC(k,.)$ and outputs $(m,t)$. Let $\Phi$ represent the set of messages that $\cal A'$ has queried.
%%    \item $\cal A$ wins if and only if (1) $Ver_{HMAC}(k,m,t)=1$ and (2) $m \notin \Phi '$.%its ancestors .
%%    \item The experiment returns 1 if the adversary wins the game, otherwise it returns 0.
%%\end{enumerate}
%%\end{game}
%Note that $Ver_{HMAC}$ is a deterministic algorithm that takes as input a key $k$, a message $m \in \{0,1\}^{*}$ and a tag $t$ and outputs bit $b$, with $b=1$  meaning $valid$ and $b=0$ meaning $invalid$.
%%Let ${\cal A'}$ be a probabilistic polynomial-time adversary and let $\varepsilon ' ($.$)$ be a function such that:
%%\begin{equation} \label{eq:3}
%%\Pr[Forge_{\cal A'}HMAC (\lambda)=1]= \varepsilon ' (\lambda)
%%\end{equation} 
%%and let $\Phi '$ represent the set of all tokens that $\cal A'$ has queried.
%Now, we define a new scheme $RAFT'=(UserIssue', ServiceIssue', Ver')$, which is similar to $RAFT$ except a random function $f(.)$ is replaced with the first HMACs in $UserIssue$ and $Ver$ algorithms (i.e., HMAC(k,p) is replaced by f(p)). \textcolor{red}{What is in brackets is actually the important part and should be upfront.}

\noindent Let $\beta(.)$ be a function so that $\Pr[ForgeUTT^0_{{\cal A},RAFT}(\lambda)=1]=\beta(\lambda)$. Since for $i=0$, $RAFT(k,p)=HMAC(k,p)$ and $RAFT'(k,p)=f(p)$ this indicates that:
\begin{align*}
    \Pr[ForgeUTT^0_{{\cal A},RAFT}(\lambda)=1]=\Pr[D^{HMAC(.)}=1]=\beta(\lambda)
\end{align*}
and 
\begin{align*}
    \Pr[ForgeUTT^0_{{\cal A},RAFT'}(\lambda)=1]=\Pr[D^{f(.)}=1]=\frac{1}{2^\lambda - C} 
\end{align*}
where $C=poly(\lambda)$ is the maximum number of queries distinguisher $D$ applies.
Therefore,
\begin{align*}
\alpha(\lambda)\geq |\Pr[D^{HMAC(.)}=1]-\Pr[D^{f(.)}=1]|\geq \beta(\lambda)-\frac{1}{2^\lambda - C}. %    \beta(\lambda)=\Pr[ForgeUTT^0_{{\cal A},RAFT}(\lambda)=1]\leq \alpha -\frac{1}{2^\lambda - C} 
\end{align*}
By the assumption that HMAC is secure (i.e. $\alpha(\lambda)$ is negligible) and because $C=poly(\lambda)$, it follows that $\beta(\lambda)$ is also negligible in $\lambda$. To be more precise:
\begin{align}
\Pr[ForgeUTT^0_{{\cal A},RAFT}(\lambda)=1]=\beta(\lambda)\leq \alpha(\lambda)+ \frac{1}{2^\lambda - C} \label{eq:raft_raftp0}
\end{align}
%\noindent It is clear that: \textcolor{red}{Spend more words here; and first define $\beta(\lambda)$ (actually the definition is incomplete -- you need to write out the game ForgingHMAC and it will have input A' which is derived from A -- again A' uses A and plays the interfaces UserIssue etc.; this does not some extra effort and work) after which you state the inequality.}
%\begin{align}
%    \Pr[ForgeUTT^0_{{\cal A},RAFT}(\lambda)=1] \leq \Pr[Forging HMAC]=\beta(\lambda) \label{eq:raft_raftp0}
%\end{align}
%\noindent $\beta(\lambda)$ is negligible if $\alpha(\lambda)$ is negligible. \textcolor{red}{This claim comes from a standard argument -- do repeat here for completeness. Again if one has an algorithm A (that is A' in the above comment), then you want to build a distinguisher for the HMAC so that you can relate $\alpha$ and $\beta$.}

\noindent Consider some adversary ${\cal A}_{i+1}$ in $ForgeUTT^{i+1}_{{\cal A}_{i+1},RAFT'}(\lambda)
    \label{eq:raft_raftp1}$. This adversary produces a token $r$ with a chain of ancestors $h_i$, where $h_0=f(p)$. We can write $r=([p,c_1,...,c_{i+1}],t)$ where $t$ is a  recursive % *known* \textcolor{red}{this is from the slides -- but rewrite and explain what this function G looks like: essentially is replays the chain} 
    function $G$ of $[p,c_1,...,c_{i+1}]$ and $f(p)$:
\begin{eqnarray*}
t&=& G([p,c_1,...,c_{i+1}],f(p)), \mbox{ where} \\
G([p,c_1],f(p)) &=& HMAC(f(p),p\parallel c_1), \mbox{ and} \\
G([p,c_1, \ldots, c_{i+1}],f(p)) &=&
HMAC(G([p,c_1,...,c_i],f(p)), p \parallel c_1 \parallel \cdots \parallel c_{i+1}).
\end{eqnarray*}
%\textcolor{blue}{\noindent G is the following recursive function:
%%$$ G([p,c_1,...,c_{i+1}],f(p))=HMAC(G([p,c_1,...,c_i],f(p)), p \parallel c_1 \parallel \cdots \parallel c_{i+1})$$
%%where $G([p,c_1],f(p))=HMAC(f(p),p \parallel c_1)$.
%\begin{align*}
% & for \quad j=0   \quad G([p,c_1],f(p))=HMAC(f(p),p \parallel c_1)\\
% & for \quad j=1   \quad G([p,c_1, c_2],f(p))=HMAC(G([p,c_1],f(p)),p \parallel c_1 \parallel c_2)\\
%& \vdots\\
%& for \quad j=i   \quad G([p,c_1,...,c_{i+1}],f(p))=HMAC(G([p,c_1,...,c_i],f(p)), p \parallel c_1 \parallel \cdots \parallel c_{i+1})
%\end{align*}}
In order to help creating a token $r$ that verifies correctly, only information about $f(p)$ can help. 
Since $f$ is a random function,$f(q)$ for $q\neq p$ does not give any information about $f(p)$, which also means that $G(.,f(p))$ and $G(.,f(q))$ are statistically independent. %\textcolor{red}{Therefore, also $G(.,f(p))$ and $G(.,f(q))$ are statistically independent.}
Thus, the information %provided by oracles
that the distinguisher $\cal D$ provides to the adversary $\cal A$ gives no useful information unless oracle $\cal O$ uses $f(p)$ for its computations.
%\textcolor{red}{this is confusing -- which oracles? let's not abuse this word to mean something else than what the oracle O meant above. -- here we mean the interface functionalities UseIssue etc.} that represent the $RAFT'$ interface give no useful information unless the oracle \textcolor{red}{here we do mean O and we mean that the oracle is queried for $p$ which output $f(p)$} uses $f(p)$ in its computations. 
We can therefore neglect all the oracle accesses that do not relate to computations with $f(p)$. This gives a new adversary  ${\cal A}_{i}$ such that:
\begin{align}
    \Pr[ForgeUTT^{i+1}_{{\cal A}_{i+1},RAFT'}(\lambda)=1]=\Pr[ForgeUTT^{i+1}_{{\cal A}_{i},RAFT'}(\lambda)=1]
    \label{eq:raft_raftp2}
\end{align}
A closer look at the structure of $RAFT$ and $RAFT'$ shows that, for $i\geq 0$,
\begin{align}
    \Pr[ForgeUTT^{i+1}_{{\cal A}_{i},RAFT'}(\lambda)=1]=\Pr[ForgeUTT^{i}_{{\cal A}_{i},RAFT}(\lambda)=1],   \label{eq:raft_raftpap}
\end{align}
which follows from the following induction argument: Let $k=f(p)$, $p'=p\parallel c_1$, and $c'_j=c_{j+1}$ for $j\in \{1,2,\ldots, i\}$. 
As a base case notice that
\begin{eqnarray*}
&& RAFT'(k,p,c_1) \\
&=&
HMAC(f(p),p\parallel c_1) \\
&=&
HMAC(k,p')\\
&=&
RAFT(k,p').
\end{eqnarray*}
Let $i\geq 1$. As an induction hypothesis assume 
$$ RAFT'(k,p,c_1,\ldots, c_i)
= RAFT(k,p',c'_1, \ldots, c'_{i-1}).$$
In exactly the same way as before we can prove the induction step
\begin{eqnarray*}
&& RAFT'(k,p,c_1,\ldots,c_{i+1}) \\
&=&
HMAC(RAFT'(k,p,c_1,\ldots,c_i), p\parallel c_1 \parallel c_2 \parallel \ldots \parallel c_{i+1}) \\
&=&
HMAC(RAFT(k,p',c'_1,\ldots,c'_{i-1}), p'\parallel c'_1 \parallel  \ldots \parallel c_{i})\\
&=&
RAFT(k,p', c'_1,\ldots,c'_i).
\end{eqnarray*}
By using induction in $i$, we conclude that the induction hypothesis holds for all $i\geq 1$, and (\ref{eq:raft_raftpap}) follows.

By defining $p_i$ and $p_i'$ as  %assume: \textcolor{red}{you mean let's define}
\begin{align}
    p_i=\Pr[ForgeUTT^{i}_{{\cal A}_{i},RAFT}(\lambda)=1] \text{ and } p_i'=\Pr[ForgeUTT^{i}_{{\cal A}_{i},RAFT'}(\lambda)=1]
    \label{eq:raft_raftp5}
\end{align}
\noindent we conclude from equations (\ref{eq:raft_raftpi}) and (\ref{eq:raft_raftp5}) 
$$|p_i-p_i'|\leq\alpha(\lambda)$$
which in turn proves
$$p_i\leq\alpha(\lambda) + p_i'.$$
\noindent From equations (\ref{eq:raft_raftp0}) and (\ref{eq:raft_raftp5}) we have
$$p_0=\beta(\lambda)$$
\noindent and from equations (\ref{eq:raft_raftpap}) and (\ref{eq:raft_raftp5}) we have:
$$p_{i+1}'=p_i.$$
%\noindent Then, \ref{eq:raft_raftpi}, \ref{eq:raft_raftp0}, and \ref{eq:raft_raftpap} give $|p_i-p_i'|\leq\alpha(\lambda)$ and $p_0\leq\beta(\lambda)$ and $p_{i+1}'=p_i$, and we have: \textcolor{red}{Separate which formulas lead to with statements about $p_i$ etc. -- give 3 separate arguments.}
These properties of $p_i$ and $p'_i$ can be combined to prove
$$p_n \leq \alpha(\lambda) +p'_n = \alpha(\lambda) + p_{n-1} \leq \ldots \leq n\cdot \alpha(\lambda) + p_0 \leq 
n\cdot \alpha(\lambda) + \beta(\lambda). $$

%\textcolor{blue}{
% $for \quad j=0:$ 
%\vspace{-.75cm}
%\begin{align*}
%  \Pr[ForgeUTT^{0}_{{\cal A}_{i},RAFT}(\lambda)=1] 
%  &= p_0\nonumber \\
%  &\leq \alpha(\lambda) + p_0' =\alpha(\lambda)+\beta(\lambda)
%\end{align*}
%$for \quad j=1$
%\vspace{-.75cm}
%\begin{align*}
%  \Pr[ForgeUTT^{1}_{{\cal A}_{i},RAFT}(\lambda)=1] 
%  &= p_1\nonumber \\
%  &\leq \alpha(\lambda) + p_1' =\alpha(\lambda) + p_0\\
%  & \leq 2 \alpha(\lambda)+\beta(\lambda)\\
%   &\vdots 
%\end{align*}
%$for \quad j=n$
%\vspace{-.75cm}
%\begin{align}
%\Pr[ForgeUTT^{n}_{{\cal A}_{i},RAFT}(\lambda)=1] 
%  &= p_n \nonumber\\
%  &\leq \alpha(\lambda) + p_n' =\alpha(\lambda) + p_{n-1}\nonumber\\
%  & \leq (n+1) \alpha(\lambda)+\beta(\lambda) \label{eq:raft_raftpFin1}
%\end{align}   }
%%\begin{align}
%%    \Pr[ForgeUTT^{n}_{{\cal A}_{i},RAFT}(\lambda)=1] &= p_n\nonumber \\
%%    &= p_n-p_n' +p_{n-1}- ... - p_1'+p_0 \nonumber\\
%%    &\leq n . \alpha(\lambda) + \beta(\lambda)
%%    \label{eq:raft_raftpFin}
%%\end{align}
We can now conclude that
\begin{align}
\Pr[ForgeUTT^{\leq n}_{{\cal A}_{i},RAFT}(\lambda)=1]&=\max_{i\leq n} \Pr[ForgeUTT^i_{{\cal A}_{i},RAFT}(\lambda)=1] \nonumber\\
& \leq \Pr[ForgeUTT^{n}_{{\cal A}_{i},RAFT}(\lambda)=1] \nonumber\\
& \leq n\cdot \alpha(\lambda)+\beta(\lambda). \label{eq:raft_raftpFin}
\end{align}
Since the adversary can at most call the oracle $poly(\lambda)$ times, $n=poly(\lambda)$. In fact, because of the policy enforcer, $n$ can not be more than the number of modules in a system.
Since $\alpha(\lambda)$ and $\beta(\lambda)$ are negligible functions in $\lambda$ and $n=poly(\lambda)$,
 (\ref{eq:raft_raftpFin}) proves the lemma.

%Since $\alpha(\lambda)$ and $\beta(\lambda)$ are negligible functions, and $n$ can not be more than the number of modules in a system, \textcolor{red}{This is an additional asusmption which needs to be added to the lemma statement itself: in fact we restrict ForgeUTT to $ForgeUTT^{\leq n}$. So, you should introduce the $ForgeUTT^i$ game and the $ForgeUTT^{\leq i}$ game *before* the lemma so that the statement of the lemma can involve this notation. In the proof here, you can write that the lemma now follows from $Pr[ForgeUTT^{\leq n}] = \max_{i\leq n} Pr[ForgeUTT^i] \leq ...$ etc.} then \ref{eq:raft_raftpFin} proves the lemma.
\end{proof}

\subsection{Proof of Lemma \ref{lem:FullRaft}}

%%%%%%%%%%%%%%%%%%%%%%%%%%%%%%%%%%%%%%%%%%
%%%%%%%%%%%%%%%  BEGIN  %%%%%%%%%%%%%%%%%%
%%%%%%%%%%%%%%%%%%%%%%%%%%%%%%%%%%%%%%%%%%
Based on Game \ref{gm:FTT} we define two other games which we will need for the proof of Lemma \ref{lem:FullRaft}:
\begin{itemize}
    \item Consider $ForgeFTT^i_{{\cal A},\Pi}$ to be the notion for forging a token with exactly $i$ commands. For this, we can define a game that is exactly as game \ref{gm:FTT} with one extra condition in Step 3, which is \textit{"m contains exactly $i$ commands"}
    \item Consider $ForgeFTT^{\leq i}_{{\cal A},\Pi}$ to be the notion for forging a token with at most $i$ commands. For this, we can define a game that is exactly as game \ref{gm:FTT} with one extra condition in Step 3, which is \textit{"m contains at most $i$ commands"}
\end{itemize}

\begin{proof}
The intuition behind the proof of this lemma is that for a PPT adversary, forging a fully tied RAFT (i.e. $\widehat{RAFT}$) without observing the input message before involves distinguishing the HMAC from a random function. 

Lets introduce the notation $\widehat{RAFT}(K,p,c_1,\ldots,c_i)$ (where $K=(k_0, k_1,\ldots, k_{n-1})$ and $n$ is the number of services) to stand for a fully tied RAFT token with $i$ commands and define $\widehat{RAFT}$ as follows:

\[
\small{\widehat{RAFT}(K,p,c_1,\ldots, c_i) =}
\begin{cases}
 HMAC(HMAC(k_0,p), p \parallel c_1) =h_1 & \small{i= 1}\\
  \resizebox{.59\hsize}{!}{$HMAC(k_{i-1}, p\parallel c_1\parallel \ldots \parallel c_i \parallel \widehat{RAFT}(K,p,c_1,\ldots,c_{i-1}))= h_i$}  & \small{n\geq i>1} \\
\end{cases}
\] \label{eq:RAFTFullEq}
Let us also simplify the descriptions of algorithms \ref{alg:GenRAFTUser}, \ref{alg:GenFRAFTService}, and \ref{alg:verRAFTFull} as follows:
\begin{itemize}
    \item For the key $k_0$ and input payload $p$ and the command $c$:
            \begin{align}
    UserIssue(k_0,p,c) = HMAC(HMAC(k_0,p), p\parallel c)= h_1=\widehat{RAFT}(K,p,c_1)\label{eq:userIssueFull}
            \end{align}
    \item For the key $k_{i-1}$ ($i \in \{2,\ldots, n\}$ where $n$ is the number of services), input token $r=(m,t)=(p\parallel c_1 \parallel \ldots \parallel c_{i-1}, h_{i-2})$, and the command $c_i$:
            \begin{align}
          ServiceIssue(i-1,K,r=(m,t), c_i) & = HMAC(k_{i-1}, m \parallel c_i \parallel t) \nonumber\\
            & = HMAC(k_{i-1}, p \parallel c_1 \parallel \cdots  \parallel c_{i-1} \parallel h_{i-1}) \nonumber\\
            & = h_i \nonumber\\
            & = \widehat{RAFT}(K,p,c_1, \ldots, c_i)           \label{eq:ServiceIssueFull}
            \end{align}
    \item For the key set $K=(k_0, k_1,\ldots, k_{n-1})$ where $n$ is the number of services, and the input token $r=(m,t)$, first, $Ver(K,r)$ decomposes $m$ and gets $\{p,c_1,\cdots,c_i\}$, then calculates: 
        \begin{align*}
            h_1 &= HMAC(HMAC(k_0,p), p \parallel c_1) \\
            \textit{Find corresponding} & \textit{ key } k_1 \textit{ for $c_2$}\\
            h_2 &= HMAC(k_1, p \parallel c_1 \parallel c_2 \parallel h_1) \\
            \textit{Find corresponding} & \textit{ key } k_2 \textit{ for $c_3$}\\
            h_3 &= HMAC(k_2, p \parallel c_1 \parallel c_2 \parallel c_3 \parallel  h_2 )\\
            \vdots\\
            \textit{Find corresponding} & \textit{ key } k_{i-1} \textit{ for $c_i$}\\
            \textit{Extract } & k_n \textit{ from $c_n$}\\
            h_i &= HMAC(k_{i-1}, p \parallel c_1 \parallel \cdots  \parallel c_i \parallel h_{i-1})
        \end{align*}
    \noindent Then, if $h_{i} = t$, returns \textit{valid}; otherwise, returns \textit{invalid}.
\end{itemize}

Let ${\cal A}$ be a probabilistic polynomial-time adversary and let $\varepsilon ($.$)$ be a function
such that:
\begin{equation} \label{eq:13}
\Pr[ForgeFTT_{{\cal A}, \widehat{RAFT}} (\lambda)=1]= \varepsilon (\lambda)
\end{equation} 
and let $\Phi$ represent the set of all tokens that $\cal A$ has queried. We want to show that $\varepsilon (\lambda)$ is negligible. 

According to the game \ref{gm:FTT}, the adversary ${\cal A}$ wins the game $ForgeFTT_{{\cal A},\widehat{RAFT}}$ if he can forge a token that is either an output of $UserIssue$ or $ServiceIssue$ for the case which the $k_i$ is not leaked. 

In other words, 
\begin{align}
\Pr[ForgeFTT_{{\cal A},\widehat{RAFT}}(\lambda) = 1] & = \Pr[ForgeFTT^{1 \leq i\leq n}_{{\cal A},\widehat{RAFT}}(\lambda) = 1] \nonumber \\
& = \max_{1\leq i\leq n}\Pr[ForgingFTT^{i}_{{\cal A}, \widehat{RAFT}}(\lambda)=1] 
\label{eq:fullyMain}
\end{align}
where $i \notin I$.

A closer look at the structure of user tied RAFT token ($RAFT$) and fully tied RAFT token ($\widehat{RAFT}$) shows that, for $i=1$,
        \begin{align}
 \widehat{RAFT}(K,p,c_1)=HMAC(HMAC(k_0,p), p \parallel c_1)=RAFT(k_0,p,c_1)  \label{eq:hatRAFT1}
        \end{align}

\noindent and for $1< i\leq n$, if we consider $k'=k_{i-1}$ and $p'=p \parallel c_1 \parallel c_2 \parallel  \ldots \parallel  c_i \parallel  h_{i-1}$ then,
        \begin{align}
         \widehat{ RAFT}(K,p,c_1,\ldots, c_i)& = HMAC(k_{i-1}, p \parallel c_1 \parallel \ldots \parallel c_i \parallel h_{i-1}) \nonumber \\
         & = HMAC(k',p') \nonumber \\
         & = RAFT(k',p') \label{eq:hatRAFTi}
        \end{align}
From lemma \ref{lem:UTRaft1} and equation (\ref{eq:hatRAFT1}) we have:
\begin{align}
\Pr[ForgingFTT^{1}_{{\cal A}, \widehat{RAFT}}(\lambda)=1]=\Pr[ForgingUTT^{1}_{{\cal A}, RAFT}(\lambda)=1]\leq  \alpha(\lambda)+\beta(\lambda) \label{eq:SemiFinalFull1}
\end{align}
where $\alpha(\lambda)$  and $\beta(\lambda)$ % is the probability that a distinguisher $\cal D$ can distinguish between the $HMAC$ function and a random function and 
are negligible functions in $\lambda$ (for more details see equation (\ref{eq:raft_raftp0})).

\noindent From lemma \ref{lem:UTRaft1} and equation (\ref{eq:hatRAFTi}) we have (for $1<i\leq n$):
\begin{align}
\Pr[ForgingFTT^{i}_{{\cal A}, \widehat{RAFT}}(\lambda)=1]=\Pr[ForgingUTT^{0}_{{\cal A}, RAFT}(\lambda)=1]\leq  \beta(\lambda) \label{eq:SemiFinalFulli}
\end{align}

From equations (\ref{eq:13}), (\ref{eq:fullyMain}), (\ref{eq:SemiFinalFull1}) and (\ref{eq:SemiFinalFulli}) we have
\begin{equation} \label{eq:15}
\varepsilon (\lambda)=\Pr[ForgeFTT_{{\cal A}, \widehat{RAFT}} (\lambda)=1]\leq \alpha(\lambda)+\beta(\lambda)
\end{equation} 

%Note that because of the policy enforces, n can not be more than the number of modules in a system. 
Since $\alpha(\lambda)$  and $\beta(\lambda)$ are negligible functions in $\lambda$, (\ref{eq:15})
proves the lemma.
\end{proof}

%% file: codes/create_vm.tex
\begin{code}
\caption{the sample command for creating a virtual machine that we used in our experiment}
\label{code:createVM}
\begin{minted}[breakanywhere]{python}
def get_create_server_sample():
    cmd = "compute/v2.1/servers"
    data ={
        "server": {
            "name": "vm_name",
            "imageRef": "ce0afaaa-e236-47c6-95e8-47c7694eb74c",
            "flavorRef": "1",
            "max_count": 1,
            "min_count": 1,
            "networks": [{"uuid": "5eeb14b4-47a9-44aa-bade-b225b7713a6b"}]
        }
    }

    cmd += " " + str(data)
    return cmd
\end{minted}
\end{code}

%% file: codes/create_vol.tex
\begin{code}
\caption{the sample command for creating a volume that we used in our experiment}
\label{code:createVol}
\begin{minted}[breakanywhere]{python}
def get_create_volume_sample():
    cmd = "volume/v2/08b72d6e4f2b465d96e9e0db2f10d232/volumes"    
    data = {
        "volume": {
            "status": "creating",
            "name": "vol_name",
            "imageRef": "ce0afaaa-e236-47c6-95e8-47c7694eb74c",
            "attach_status": "detached",
            "volume_type": "lvmdriver-1",
            "size": 1
        }
    }
    cmd += " " + str(data)
    return cmd
\end{minted}
\end{code}

%% file: codes/howToExtractPayload.tex
\begin{code}
\caption{The python code for extracting the payload of a project scoped token given a Fernet key and token.}
\label{code:extractingPayload}
\scriptsize
\begin{minted}[breakanywhere]{python}
# this code is adapted from the source code of Keystone Stein
from cryptography.fernet import Fernet
import msgpack
import uuid

def restore_padding(token):
    mod_returned = len(token) % 4
    if mod_returned:
        missing_padding = 4 - mod_returned
        token += '=' * missing_padding
    return token

key = "Qh4ZzunoX36Ri0TKVa3bXqzTQKzwqT3G4JfmGw1ZNtU="
f = Fernet(key)
token="gAAAAABdpxhmvMe_byl3qKlJ0KVXizdSyL_38Idxam2ap7O1T9_xzX9eVJ6WCozRKlXjH6oZlDuOyS0nI_57u0G0ceOt7coUtDPPI1TipydgxMekVNtbhdHuR8A9BMvY1pPAVkGV_23Hd_Ste0eiTXP7m_7W77Vj3X2qGkjkeuinyGZsTclYZOc"
print("Token Length = %d" % len(token)) 
token = restore_padding(token)
serialized_payload = f.decrypt(token.encode('utf-8'))
print("Payload Length = %d" %len(serialized_payload))

versioned_payload = msgpack.unpackb(serialized_payload)
version, payload = versioned_payload[0], versioned_payload[1:]
print("version = %d" % version) # prints 2 which means Project Scoped Token

(is_stored_as_bytes, user_id) = payload[0]
if is_stored_as_bytes:
    user_id = uuid.UUID(bytes=user_id)
print("User Id = %s" % user_id)

print("Method = %d" % payload[1]) # prints 2 which means "Password"

(is_stored_as_bytes, project_id)=payload[2]
if is_stored_as_bytes:
    project_id = uuid.UUID(bytes=project_id)

print("Project Id = %s" % project_id)

print("Expiration Time %s " % payload[3])
print("Audit Id = %s"% payload[4])
\end{minted}
\end{code}

%% file: codes/RaftLib.tex
\begin{code}
\caption{The source code of user tied RAFT Library.}
\label{code:RAFTLib}
\scriptsize
\begin{minted}[breakanywhere]{python}
#importing the required library
import os  
import time
import six

from cryptography.hazmat.backends import default_backend
import base64
from cryptography import fernet
from cryptography.hazmat.primitives.hmac import HMAC
from cryptography.hazmat.primitives import hashes
from cryptography import utils
from struct import pack,unpack, unpack_from

#KeystoneRAFT is developed to be added to Keystone. It allows Keystone to verfy RAFT tokens
class KeystoneRAFT(object):  
    expirationTime = 10
    @classmethod
    def restore_padding(cls, token):
        """Restore padding based on token size.

        :param token: token to restore padding on
        :type token: six.text_type
        :returns: token with correct padding

        """
        # Re-inflate the padding
        mod_returned = len(token) % 4
        if mod_returned:
            missing_padding = 4 - mod_returned
            token += '=' * missing_padding
        return token

    #the function which is used for initialization of new object
    def __init__(self, backend=None):
        if backend is None:
            backend = default_backend()
        self._backend = backend

    def isRAFT(self, token):
        """
            Checks the first byte of the given token to find out if it is a RAFT or NOT.
            Returns True if the 
        """
        if six.indexbytes(base64.urlsafe_b64decode(ClientRAFT.restore_padding(token)),0)== 0x91:
            return True
        else:
            return False

    def CheckExpirationTime(self,token,pos):
        """
            Check the expiration time of a token adn rise an exception if the token is already expired
        """
        exp_time, = unpack_from(">Q",token,pos)
        if self.expirationTime > exp_time:
            self.expirationTime =exp_time

        if exp_time < time.time():
            raise Exception("the token has expired")
        return exp_time
        
    def GetKey(self, token):
        """
            This function recursively extracts the parent messages and checks their expiration time until it gets to the root token. 
            Then it starts to recalculate the hmac of every parent's messages.  
        """
        v, = unpack_from(">B",token,0)
        if v==0x91 :
            #it must be a RAFT
            lenId, = unpack_from(">H",token,1)
            OriginalId, = unpack_from(">" + str(lenId)+"s", token,3)
            cmd = token[19+lenId:]
            self.CheckExpirationTime(token, 3+lenId)
            keys, cmds = self.GetKey(OriginalId)
            cmds.append(cmd)
            sign_key = keys[:16]
            h = HMAC(sign_key, hashes.SHA256(), self._backend)
            h.update(token)  
            hmac=h.finalize()
            return hmac, cmds
        else:
            # when we get the root token which is supposed to be Fernet token
            sign_key = base64.urlsafe_b64decode(self.SourceKey)[-32:-16]
            h = HMAC(sign_key, hashes.SHA256(), self._backend)
            h.update(token)
            signature = h.finalize()
            self.fernetToken = base64.urlsafe_b64encode(token+signature)
            return signature,[]
        return ""

    def ValidateRAFT(self, token):
        """
            This function get a token and using GetKey function rebulds the HMAC of the message part of the token. 
            If calculated HMAC fits with the token signature, then the token is valid; Otherwise the function raises an exception error
        """
        token = self.restore_padding(token)
        token = base64.urlsafe_b64decode(token)
        v, = unpack_from(">B",token,0)
        self.SourceKey = 'Qh4ZzunoX36Ri0TKVa3bXqzTQKzwqT3G4JfmGw1ZNtU='
#        self.SourceKey = fernet_utils.load_keys()
        if v==0x91 :
            lenId, = unpack_from(">H",token,1)
            OriginalId, =  unpack_from(">"+str(lenId)+"s",token,3)
            self.expirationTime = self.CheckExpirationTime(token, 3+lenId)
            cmd = token[19+lenId:-32] 
            print(cmd)
            keys,cmds = self.GetKey(OriginalId)
            cmds.append(cmd)
            sign_key = keys[:16]
            h = HMAC(sign_key, hashes.SHA256(), self._backend)
            h.update(token[:-32])  
            hmac=h.finalize()
            if hmac== token[-32:]:
                #print("It is valid")
                return True,cmds
            else:
                raise Exception("Error: Not a RAFT token (MAC)")
        else:
            raise Exception("Error: Not a RAFT token (Format)")
        return ""

#ClientRaft is used for creating a new RAFT from either a Fernet or other RAFT.
class ClientRAFT(object):  
    _lifeTime = 1  #feault lifetime for a RAFT
#the function which is used for initialization of new object
    def __init__(self, parent_token=None, backend=None):
        if backend is None:
            backend = default_backend()
        self._backend = backend

        if parent_token is not None:            
            self.SetParentToken(parent_token)

    @classmethod
    def restore_padding(cls, token):
        """Restore padding based on token size.

        :param token: token to restore padding on
        :type token: six.text_type
        :returns: token with correct padding

        """
        # Re-inflate the padding
        mod_returned = len(token) % 4
        if mod_returned:
            missing_padding = 4 - mod_returned
            token += '=' * missing_padding
        return token

    def SetParentToken(self,parent_token):
        """
            Allows to define/redfine the parent token of this object 
        """
        parent_token= self.restore_padding(parent_token)
        self._parentToken = base64.urlsafe_b64decode(parent_token)
        #the last 32 bytes of the parent token is the HMAC, which the first 16 bytes of it is the signing of current token
        self._signKey = self._parentToken[-32:-16]
        self._encryptionKey = self._parentToken[-16:]
        self._id = self._parentToken[:-32]

    def SetCommand(self, command):
        """
            Specifinying a command that is supposed to be added to the parent token
        """
        self._command = command

    def Finalize(self, life_time = None):
        """
        Packs all the items needed to be in the token, 
            specifies the lifetime,
            calculates the HMAC of the packed message,
            and convert the result to a base64 string, which is the new RAFT token
        """ 
        if life_time is not None:
            self._lifeTime = life_time

        self._NewToken = pack(">BH" + 
                              str(len(self._id))+"s",
                              0x91,
                              len(self._id),
                              self._id)
        self._command=pack(">Q8s"+str(len(self._command))+"s", int(time.time())+self._lifeTime,os.urandom(8),self._command) #bytearray(self._command,"utf8")
        self._NewToken+=self._command
        self.h = HMAC(self._signKey, hashes.SHA256(), self._backend)
        self.h.update(self._NewToken)
        self._NewToken += self.h.finalize()
        return base64.urlsafe_b64encode(self._NewToken).rstrip(b'=')

\end{minted}
\end{code}        

%% file: codes/validateToken.tex
\begin{code}
\caption{The validate\_token function of the TokenFormatter class after applying the required changes.}
\label{code:validateToken}
\scriptsize
\begin{minted}[breakanywhere]{python}
    def validate_token(self, token):
# The following code is added to check if the token is RAFT, then process the RAFT using our library
        try:
            raftFlag = False
            kr = RAFT.KeystoneRAFT()
            if kr.isRAFT(token):
                raftFlag = True
                kr.ValidateRAFT(token)
                token = kr.fernetToken
         except Exception as ex:
            template = "An exception of type {0} occurred. Arguments:\n{1!r}"
            message = template.format(type(ex).__name__, ex.args)
#End of processing a RAFT.

#The old code for processing Fernet Token
        serialized_payload = self.unpack(token)
        versioned_payload = msgpack.unpackb(serialized_payload)
        version, payload = versioned_payload[0], versioned_payload[1:]        
        for payload_class in _PAYLOAD_CLASSES:
            if version == payload_class.version:
                (user_id, methods, system, project_id, domain_id,
                 expires_at, audit_ids, trust_id, federated_group_ids,
                 identity_provider_id, protocol_id, access_token_id,
                 app_cred_id) = payload_class.disassemble(payload)
                break
        else:
            raise exception.ValidationError(_(
                'This is not a recognized Fernet payload version: %s') %
                version)

        if isinstance(system, bytes):
            system = system.decode('utf-8')

        issued_at = TokenFormatter.creation_time(token)
        issued_at = ks_utils.isotime(at=issued_at, subsecond=True)

# The following code, adjusts the expiration time if the token was RAFT

        if raftFlag:            
            expires_at = kr.expirationTime
            expires_at = BasePayload._convert_float_to_time_string(expires_at)         

        expires_at = timeutils.parse_isotime(expires_at)
        expires_at = ks_utils.isotime(at=expires_at, subsecond=True)
# end of expiration adjustment

        return (user_id, methods, audit_ids, system, domain_id, project_id,
                trust_id, federated_group_ids, identity_provider_id,
                protocol_id, access_token_id, app_cred_id, issued_at,
                expires_at)
\end{minted}
\end{code}

%% file: codes/expSet1.tex
\begin{code}
\caption{The source code of the first set of our experiments.}
\label{code:firstExp}
\scriptsize
\begin{minted}[breakanywhere]{python}
import requests
import os
import time
import json
from RAFT import ClientRAFT,KeystoneRAFT

base_url = "http://192.168.122.29" #The IP address of our OpenStack server

def get_scoped_token(username, password, domain,project_name):
    data= { 
        "auth": { 
            "identity": { 
                "methods": ["password"],
                "password": {
                    "user": {
                        "domain": {
                            "name": domain
                        },
                        "name": username,
                        "password": password
                    }
                }
            },
            "scope": { 
                "project": {
                    "domain": { 
                        "name": "Default" 
                    },
                    "name":  project_name
                } 
            }
        }
    }
    url = base_url + "/identity/v3/auth/tokens"
    r = requests.post(url,json=data)
    return r.headers["X-Subject-Token"]

def check_token(token):
    url = base_url + "/identity/v3/auth/tokens"
    h = {"X-Auth-Token" : token,"X-Subject-Token" : token}
    r = requests.get(url, headers=h)
    return r

def firstExp1():
    print("Creating 100 Fernet tokens")
    start_time= time.time()
    for i in range(0,100):
        fernet_token = get_scoped_token("admin","123","Default","admin")
    execution_time = time.time() - start_time
    print(execution_time)

def firstExp2():
    print("Creating 100 RAFTs + a Fernet token")
    start_time= time.time()
    fernet_token = get_scoped_token("admin","123","Default","admin")
    for i in range(0,100):
        raft_builder = ClientRAFT(fernet_token)
        raft_builder.SetCommand("")
        raft_builder.Finalize()        
    execution_time = time.time() - start_time
    print(execution_time)

def firstExp3():
    print("Creating and validating 100 Fernet tokens")
    start_time= time.time()
    for i in range(0,100):        
        fernet_token = get_scoped_token("admin","123","Default","admin")
        check_token(fernet_token)
    execution_time = time.time() - start_time
    print(execution_time)

def firstExp4():
    print("Creating and validating 100 RAFTs with no command")
    start_time= time.time()
    fernet_token = get_scoped_token("admin","123","Default","admin")
    for i in range(0,100):
        raft_builder = ClientRAFT(fernet_token)
        raft_builder.SetCommand("")
        raft_token= raft_builder.Finalize(1)
        check_token(raft_token)
    execution_time = time.time() - start_time
    print(execution_time)

def firstExp5to7(CmdLength):
    print("Creating and validating 100 RAFTs with a hypothetical 200 characters command")
    start_time= time.time()
    fernet_token = get_scoped_token("admin","123","Default","admin")
    cmd = "a" * CmdLength
    for i in range(0,100):
        raft_builder = ClientRAFT(fernet_token)
        raft_builder.SetCommand(cmd)
        raft_token= raft_builder.Finalize(10)
        check_token(raft_token)
    execution_time = time.time() - start_time
    print(execution_time)

def firstExpIssueTime():
    print("Locally Creating 100 RAFTs experiment")
    fernet_token = get_scoped_token("admin","123","Default","admin")
    cmd = "a" * 20
    start_time= time.time()
    for i in range(0,100):
        raft_builder = ClientRAFT(fernet_token)
        raft_builder.SetCommand(cmd)
        raft_token= raft_builder.Finalize(10)
    execution_time = time.time() - start_time
    print(execution_time)

def firstExpVerificationTime():
    print("Locally creating and validation 100 RAFTs experiment")
    fernet_token = get_scoped_token("admin","123","Default","admin")
    cmd = "a" * 20
    total_time = 0
    for i in range(0,100):
        raft_builder = ClientRAFT(fernet_token)
        raft_builder.SetCommand(cmd)
        raft_token= raft_builder.Finalize(10)
        start_time= time.time()
        kv = KeystoneRAFT();
        kv.ValidateRAFT(raft_token)
        execution_time = time.time() - start_time
        total_time+=execution_time
    print(total_time)
\end{minted}
\end{code}

%% file: codes/expSet2.tex
\begin{code}
\caption{Examining the overhead of using RAFT with some commands}
\label{code:secExp}
\scriptsize
\begin{minted}[breakanywhere]{python}
import requests
import os
import time
import json
import expCommandLength
from RAFT import ClientRAFT,KeystoneRAFT

base_url = "http://192.168.122.29" # the IP address of our OpenStack server

def get_project_list(token):
    url = base_url + "/identity/v3/projects"
    h = {"X-Auth-Token" : token,"X-RAFT-Token" : token}
    r = requests.get(url,headers=h)    
    if r.status_code == 200:
        return r.content
    else:
        print(r.content)
        return None

def create_VM(token, vm_name):        
    url = base_url + "/compute/v2.1/servers"
    h = {"X-Auth-Token" : token,"X-RAFT-Token" : token, "Accept": "application/json", "Content-Type": "application/json", "User-Agent": "python-novaclient"}
    data ={
        "server": {
            "name": vm_name,
            "imageRef": "ce0afaaa-e236-47c6-95e8-47c7694eb74c",
            "flavorRef": "1",
            "max_count": 1,
            "min_count": 1,
            "networks": [{"uuid": "5eeb14b4-47a9-44aa-bade-b225b7713a6b"}]
        }
    }
    r = requests.post(url,json=data,headers=h)
    return r

# the default values used in the following function are only valid in our server. 
def create_Volume(token, volume_name, vol_size=1 , project_id="08b72d6e4f2b465d96e9e0db2f10d232" , imageId= "ce0afaaa-e236-47c6-95e8-47c7694eb74c" ):        
    url = base_url + "/volume/v2/" + project_id +  "/volumes"    
    h = {"X-Auth-Token" : token,"X-RAFT-Token" : token, "Accept": "application/json", "Content-Type": "application/json", "User-Agent": "python-novaclient"}
    data = {
        "volume": {
            "status": "creating",
            "name": volume_name,
            "imageRef": imageId,
            "attach_status": "detached",
            "volume_type": "lvmdriver-1",
            "size": vol_size
        }
    }
        
    r = requests.post(url,json=data,headers=h)
    return r

def get_imagelist(token):
    url = base_url + "/image/v2/images"
    h = {"X-Auth-Token" : token,"X-RAFT-Token" : token}
    r = requests.get(url,headers=h)
    if r.status_code == 200:
        j = json.loads(r.content)
        return j
    else:
        print(r.content)
        return None

def SecondExp1():
    print("Creating a Volume using Fernet")
    for j in range(0,5):
        fernet_token = get_scoped_token("admin","123","Default","admin")
        start_time= time.time()    
        for i in range(0,1):
            create_Volume(fernet_token, "vreza"+str(i))
        execution_time = time.time() - start_time
        print(execution_time)

def SecondExp2():
    print("Creating a Volume using RAFT")
    for j in range(0,5):
        fernet_token = get_scoped_token("admin","123","Default","admin")
        start_time= time.time()    
        for i in range(0,1):
            raft_builder = ClientRAFT(fernet_token)
            raft_builder.SetCommand(expCommandLength.get_create_volume_sample())
            raft_token= raft_builder.Finalize(100)
            create_Volume(raft_token, "vreza"+str(i))
        execution_time = time.time() - start_time
        print(execution_time)

def SecondExp3():
    print("Creating a Virtual Machine using Fernet")
    for j in range(0,5):
        fernet_token = get_scoped_token("admin","123","Default","admin")
        start_time= time.time()    
        for i in range(0,1):
            create_VM(fernet_token, "vm"+str(j))
        execution_time = time.time() - start_time
        print(execution_time)

def SecondExp4():
    print("Creating a Virtual Machine using RAFT")
    for j in range(0,5):
        fernet_token = get_scoped_token("admin","123","Default","admin")
        start_time= time.time()    
        for i in range(0,1):
            raft_builder = ClientRAFT(fernet_token)
            raft_builder.SetCommand(expCommandLength.get_create_server_sample())
            raft_token= raft_builder.Finalize(100)
            create_VM(raft_token, "vm"+str(j))
        execution_time = time.time() - start_time
        print(execution_time)

def SecondExp5():
    print("Getting image list using Fernet")
    for j in range(0,5):
        fernet_token = get_scoped_token("admin","123","Default","admin")
        start_time= time.time()    
        for i in range(0,1):
            get_imagelist(fernet_token)
        execution_time = time.time() - start_time
        print(execution_time)

def SecondExp6():
    print("Getting image list using RAFT")
    for j in range(0,5):
        fernet_token = get_scoped_token("admin","123","Default","admin")
        start_time= time.time()    
        for i in range(0,1):
            raft_builder = ClientRAFT(fernet_token)
            raft_builder.SetCommand(expCommandLength.get_image_list_command())
            raft_token= raft_builder.Finalize(2)
            get_imagelist(raft_token)
        execution_time = time.time() - start_time
        print(execution_time)

def SecondExp7():
    print("Getting project list using Fernet")
    for j in range(0,5):
        fernet_token = get_scoped_token("admin","123","Default","admin")
        start_time= time.time()    
        for i in range(0,1):
            get_project_list(fernet_token)
        execution_time = time.time() - start_time
        print(execution_time)

def SecondExp8():
    print("Getting project list using RAFT")
    for j in range(0,5):
        fernet_token = get_scoped_token("admin","123","Default","admin")
        start_time= time.time()    
        for i in range(0,1):
            raft_builder = ClientRAFT(fernet_token)
            raft_builder.SetCommand(expCommandLength.get_project_command())
            raft_token= raft_builder.Finalize(2)
            get_project_list(raft_token)
        execution_time = time.time() - start_time
        print(execution_time)

\end{minted}
\end{code}